\documentclass[%
prl, aps,amsmath,amssymb,
reprint,%
superscriptaddress
]{revtex4-2}
\usepackage{dcolumn} 
\usepackage{multirow}
\usepackage[T1]{fontenc}
\usepackage{dsfont}
\usepackage{verbatim}
\usepackage{bm}
\usepackage{dblfloatfix}
\usepackage{hyperref} 
\usepackage{cleveref} 
\usepackage{floatrow}  
\usepackage{natbib}
\usepackage{graphicx}
\usepackage{color}
\usepackage[normalem]{ulem}
\usepackage[usenames,dvipsnames]{xcolor}
\usepackage[caption=false]{subfig}
\usepackage{graphicx}

\floatstyle{plaintop}
\restylefloat{table}

\usepackage{colortbl}
\usepackage{makecell}%To keep spacing of text in tables
\usepackage{hhline}
\usepackage{multirow}
\usepackage{xr}
\newcommand{\icomplex}{\dot\iota}%other notation: \dot\iota, \imath, \dot\imath, {\mathrm{i}}

\begin{document}

\preprint{APS/123-QED}

\title{Non-Markovian modeling of non-equilibrium fluctuations and dissipation in active viscoelastic biomatter}

\author{Amir Abbasi}
\affiliation{
	Fachbereich Physik, Freie Universit\"at Berlin, Arnimallee 14, 14195 Berlin, Germany}
\author{Roland R. Netz}
\affiliation{
	Fachbereich Physik, Freie Universit\"at Berlin, Arnimallee 14, 14195 Berlin, Germany}%
\author{Ali Naji}
\thanks{a.naji@ipm.ir}
\affiliation{School of Nano Science, Institute for Research in Fundamental Sciences (IPM), Tehran 19395-5531, Iran}

%\date{\today}

\begin{abstract}
Based on a  Hamiltonian that incorporates the elastic coupling between a tracer  and 
active  particles, 
we derive a   generalized Langevin model  
for the non-equilibrium mechanical response of active viscoelastic biomatter.
Our model accounts for the  power-law viscoelastic response of the embedding polymeric network
as well as for the non-equilibrium energy transfer between active and tracer particles. 
Our analytical expressions for the frequency-dependent
response function and the positional autocorrelation function agree nicely with
experimental data for red blood cells and actomyosin networks with and without ATP.
The fitted  effective active-particle  temperature, 
elastic constants and effective friction coefficients of our model allow straightforward physical interpretation.
\end{abstract}

%\keywords{}
\maketitle
%%%%%%%%%%%%%%%%%%%%%%%%%%%%%%%%%%%%%%%%%%%%%
%%%%%%%%%%%%%%%%%%%%%%%%%%%%%%%%%%%%%%%%%%%%%
%\section{Introduction}

Viscoelastic gels  such as permanently or transiently cross-linked networks of semiflexible polymers are important soft  biological materials  \cite{Schnurr1997Determining,Joanny2009active,Osada1998}. 
The polymeric nature of such  gels is responsible for their salient rheological properties, 
including their frequency-dependent response to external forces. 
The cell cytoskeleton is a prime example of a viscoelastic gel \cite{Bershadsky2012,Elson1988}
and determines cell shape, motility and division \cite{Juelicher2007,Fletcher2010}.
Even though the cytoskeleton has a complex structure, its mechanical properties at small deformations are determined mainly by the 
underlying F-actin network \cite{Lieleg2010structure,Pelletier2003structure}.
Even in equilibrium,  the dynamic properties of  F-actin are complex and interesting \cite{Hiraiwa2018systematic}. 
In vivo,  constant F-actin (de)polymerization and  interactions with  myosin motors  consume naturally occurring adenosine-tri-phosphate (ATP)  
and drive the viscoelastic network  into a non-equilibrium (NEQ) state \cite{Rayment1993structure,Martin2009pulsed,Mizuno2007nonequilibrium},
where the  motors act as active cross-links and produce local tension \cite{Joanny2009active}.  

Another illustrious example of active viscoelastic biomatter are red blood cells (RBCs), 
whose structural components include a filamentous  protein (spectrin) scaffold underneath the enclosing cell membrane \cite{Mills2004nonlinear,Puig-de-Morales-Marinkovic2006viscoelasticity,Tuvia1997cell,Faris2009membrane,Betz2012time}. 
%The spectrin network is anchored to the cell membrane at junctional protein complexes, that form a triangular cytoskeletal grid. 
The mechanical response and spectra of membrane fluctuations (RBC flickering) display active NEQ signatures that are attributed to the membrane ion-pump activity and the ATP-induced reorganization of the spectrin network  \cite{Girard2005passive,Turlier2016,Bernheim2018}.  
% a plasma membrane proteins which are in general responsible for cell signaling, ions inlet/outlet regulations, etc. For example, band-3 proteins are responsible for uptake regulation of toxicants like $CO_2$ in Red Blood Cells (RBCs).

Supported by high-resolution experimental data \cite{Radler1995fluctuation,Mizuno2007nonequilibrium,Manneville1999,Mizuno2008,Turlier2016,Betz2009,Park2010metabolic,Betz2012time}, 
a key line of inquiry has been to examine  violations of the fluctuation-dissipation theorem (FDT)
in these NEQ systems. 
The FDT holds for near-equilibrium (near-EQ) dynamical processes, but not for
 active NEQ processes, as indeed corroborated by microrheology experiments \cite{Mizuno2007nonequilibrium,Turlier2016}. 
 Typical experiments involve a micron-sized tracer bead 
immersed in the active medium whose motion is tracked and   which is either allowed to move spontaneously or is
driven by a  laser beam (passive versus active microrheology). 
Active microrheology allows to simultaneously measure the  mechanical-response  and the positional autocorrelation function
and thereby to check the validity of the FDT.

Theoretical studies of FDT violation focused on externally forced systems \cite{Mauri2006violation,Berthier2002nonequilibrium,Szamel2014self}, glasses \cite{Grigera1999observation} and active systems \cite{Mizuno2007nonequilibrium,Turlier2016,Fodor2016how}. In a generalized approach based on the Langevin equation, NEQ fluctuations have been modeled by an additive athermal noise acting on particles in a viscous fluid \cite{Bernheim2018,Ben-Isaac2011effective}. A similar strategy has been used to model active dynamical undulations of elastic biological membranes \cite{Gov2004cytoskeleton,Bernheim2018,Rodriguez2015,Gov2005red} and several computational models have explicitly incorporated active pumps \cite{Turlier2016,Lin2006nonequilibrium}.
A non-equilibrium multicomponent elastic model
for the interaction of motors and a tracer particle was recently shown to describe the frequency-dependence of the experimentally observed 
FDT violation \cite{Mizuno2007nonequilibrium} rather well  \cite{Netz2018}. 
%it is shown that a single-parameter quantifies deviations from equilibrium in two distinct manifestations: Departures of the stationary-state, phase-space probability weight from the Boltzmann distribution, and the violation of the standard fluctuation-dissipation theorem. 

In this paper, 
%highlighting the FDT violation approach, 
we model  the non-equilibrium mechanical response in active biomatter by taking into account the medium viscoelasticity
and the elastic coupling between  tracer and the active particles.
We derive analytical expressions for mechanical-response and spatial correlation functions, by simultaneously 
fitting these functions to experimental  microrheological data for RBCs \cite{Turlier2016} and reconstituted  actomyosin networks \cite{Mizuno2007nonequilibrium} we extract all parameters describing the viscoelastic medium and the elastic coupling
between tracer and active particles. 
We find that power-law viscoelasticity  describes the experimental data better than the 
standard Maxwell and Kelvin-Voigt viscoelastic models.
The extracted viscoelastic exponent is close to $\kappa=3/4$ for actomyosin networks and close to unity
for RBC, in agreement with previous studies. Interestingly,
in the presence of ATP
the effective temperature ratio between active and tracer particles  is above 20 for both systems,
demonstrating that these systems are very far from equilibrium. 
Our analysis and the quantitative agreement established with  experiments 
demonstrates the generic capability of our multicomponent elastic model  and our hybrid approach that combines  relevant 
particle-based and hydrodynamic aspects of the problem into a single framework and enables closed-form predictions.

\squeezetable
\begin{table*}
	\begin{ruledtabular}
		\caption{Model parameters obtained from fits to the experimental data}\vskip-3mm
		\begin{tabular}{lcc|ccccc}
			\textrm{Exp.}&
			\textrm{$\kappa$}&
			\textrm{$\gamma[\text{pN.s}^\kappa/\mu\text{m}]$}&
			\textrm{$K[\text{pN}/\mu\text{m}]$}&
			\textrm{$\hat{k}[\text{pN}/\mu\text{m}]$}&
			\textrm{$\hat{\gamma}_\text{a}[\text{pN.s}^\kappa/\mu\text{m}]$}&
			\textrm{$\alpha$}&
			\textrm{}\\ 
			\colrule
			RBC:&&&&& \\
			$\;\;$EQ  &$0.926\pm0.0655$ & $0.120\pm0.00295$ & $5.51\pm0.109$ & $1.31\pm0.137$ & $0.257\pm0.0219$ & $0.985\pm0.197$ &\\ 
			$\;\;$NEQ  &$0.939\pm0.0596$ & $0.0749\pm0.00164$ & $1.66\pm0.0289$ & $0.283\pm0.0361$ & $0.155\pm0.0186$ & $36.3\pm4.76$\\
			ACM:&&&&& \\
			$\;\;$EQ  &$0.737\pm0.0388$ & $6.053\pm0.785$ & $90.5\pm4.30$ & $40.3\pm4.28$ & $41.1\pm7.26$ & $-0.035\pm0.144$ &\\
			$\;\;$NEQ  &$0.731\pm0.0616$ & $7.082\pm0.891$ & $189.0\pm8.65$ & $98.5\pm13.6$ & $61.9\pm6.07$ & $23.37\pm2.83$ & \\
		\end{tabular}\label{table:fitparams}
	\end{ruledtabular}
\end{table*}
%%%%%%%%%%%%%%%%%%%%%%%%%%%%%%%%%%%%%%%%%%%%%%%%%%%%%%%%%%%%%%%%%%%%%%%%%%%%%%%%%%%%%%%%%%%%%%%%%%
\begin{figure*}
	\caption{{\em RBC flickering.} Comparison of  experimental data (symbols) \cite{Turlier2016} and our
		model predictions for the positional autocorrelation function $\omega \tilde{C}(\omega)/(2k_{\text{B}}T)$
		in Eq.~(\ref{eq:powerlawcorr}) (solid lines) 
		and the imaginary part of the response function $\tilde{\chi}''(\omega)$  Eq.~(\ref{eq:powerlawchi}) (broken lines) 
		for  (a) the EQ scenario, i.e. ATP-depleted RBCs, and (b) the NEQ scenario, i.e. RBCs in the presence of ATP. 
		In the insets, the spectral function $\Xi(\omega)$ from the experiments  (symbols) is compared with our analytical prediction
		Eq.~(\ref{eq:powerlawspec}).
		(c) Comparison of the real and   imaginary parts of the response function $\tilde{\chi}'(\omega)$  and $\tilde{\chi}''(\omega)$ for the NEQ and EQ scenarios, indicating a 
		larger response (i.e. a softening and less dissipation) 
		in the NEQ case in particular at low frequencies.
		The dotted lines in all subfigures  indicate the model prediction for the real part of the response function $\tilde{\chi}'(\omega)$ (see SI). }
	\label{fig:RBC}
	\includegraphics[width=\linewidth]{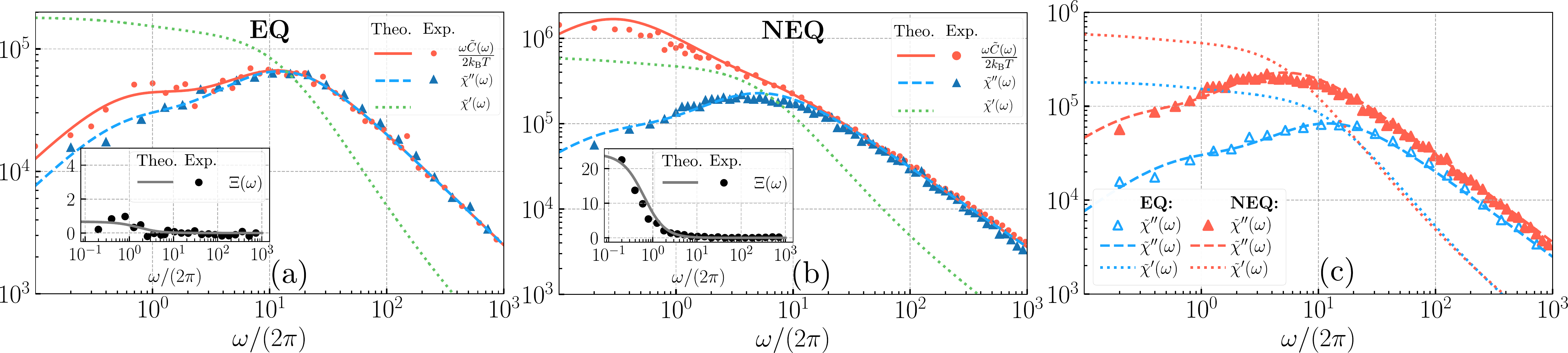}
	\vskip-2mm
\end{figure*}
%\section{\label{sec:model}Model and framework}

Inspired by typical microrheology experiments  \cite{Mizuno2008,Mizuno2007nonequilibrium,Turlier2016}, 
we assume a single tracer bead of mass $M$, embedded in an isotropic viscoelastic network
and coupled elastically via harmonic springs of strength $k$ with $n$ active particles of mass $m$. 
The tracer is trapped in an external harmonic potential of elastic constant $K$. 
The Hamiltonian reads 
\begin{equation}
	H=\frac{K}{2}X^{2}+\frac{M}{2} \dot{X}^{2}+\sum_{i=1}^{n}\frac{k}{2}(X-x_{i})^{2}+\sum_{i=1}^{n}\frac{m}{2} \dot{x}_i^{2},
	\label{Hamiltonian}
\end{equation} 
where $X$ is the position of the tracer 
and $x_i$ are the   active-particle positions ($i=1,\cdots\,,n$),  
$ \dot{X}$ and $ \dot{x}_i$ are the tracer and active-particle velocities. 
We assume that the system is  isotropic and therefore  use one-dimensional displacement variables,
 the extension to multidimensional variables is straightforward \cite{Schwarz2013,Netz2018}.  
The system dynamics is assumed to follow a generalized Langevin equation (GLE), 
which  in addition to the deterministic (Hamiltonian) forces, 
contains thermal and active random forces as well as  time-dependent friction forces that describe 
the polymeric network viscoelasticity. 
The latter is modeled using a general memory kernel $\mathcal{K} (t)$ \cite{Goychuk2012,Mainardi1997}, 
yielding the GLEs
\begin{subequations}\label{systemeq}
	\begin{align}
%		\dot{X}(t)&=V(t),\\		
		M\ddot{X}(t)=&-\gamma\int_{-\infty}^{t} \! {\mathrm{d}}t' \, \mathcal{K}_+(t-t')\dot{X}(t') \;-(nk+K)X(t)\nonumber\\&+k\sum_{i}x_{i}(t)+ F(t),\\
%		\dot{x}_{i}(t)&=v_{i}(t),\\		
		m\ddot{x}_{i}(t)=&-\gamma_\text{a}\int_{-\infty}^{t}\!{\mathrm{d}}t'\, \mathcal{K}_+(t-t')\dot{x}_i(t') \;-k(x_{i}(t)-X(t))\nonumber\\&+ f_i(t).\label{momega}
	\end{align}
\end{subequations}
Here, $\gamma$ and $\gamma_\text{a}$ are  drag coefficients 
and $F(t)$ and $f_i(t)$ are random forces 
acting on the tracer and active particles, respectively, 
with zero mean and second moments given by 
$ \langle F(t)F(t')\rangle=k_\text{B}T\gamma\mathcal{K}_+(|t-t'|)$ and 
$ \langle f_i(t)f_j(t')\rangle=k_\text{B}T_\text{a}\gamma_\text{a}\delta_{ij}\mathcal{K}_+(|t-t'|)$.
In NEQ the temperatures characterizing the random forces acting on the active and tracer particles, 
$T_\text{a}$ and $T$,  are unequal and the NEQ parameter
$\alpha = T_\text{a}/T - 1$ becomes non-zero \cite{Netz2018}.
Guided by previous results for biological membranes and polymeric networks \cite{Fabry2001scaling,Bursac2007cytoskeleton,Costa2008,Mizuno2007nonequilibrium} and theoretical works \cite{Kappler2019cyclization,Taloni2010generalized}, 
we use  a normalized power-law memory kernel,  $\mathcal{K}_+(t)=t^{-\kappa}/\Gamma(1-\kappa)$,  
where $\Gamma(x)$ is the Gamma function. 

Note that by construction,  the tracer particle and the active particles themselves obey the FDT since the 
random force correlation equals the friction kernel $\mathcal{K}_+(t)$, NEQ is therefore  introduced via the coupling
of tracer and active particles. Also, the friction kernel of active and tracer particles is assumed to be identical, 
which reflects that both particles are assumed to be coupled to the same polymeric network.

\begin{figure*}[b!]
	\caption{{\em Actomyosin networks.} In analogy to Fig.~\ref{fig:RBC}, here we compare our model predictions 
		for  $\omega \tilde{C}(\omega)/(2k_{\text{B}}T)$ and  $\tilde{\chi}''(\omega)$  (solid and broken curves) 
		with experimental actomyosin-network data from Ref.  \cite{Mizuno2007nonequilibrium} in the (a) absence  and (b) presence of ATP. 
		The insets show the spectral function $\Xi(\omega)$ defined in Eq.  \eqref{spectral}.  
		(c) Actomyosin networks exhibit a decrease of the real and imaginary response  $\tilde{\chi}'(\omega)$ and  $\tilde{\chi}''(\omega)$ in NEQ.}
	\label{fig:ACM}
	\includegraphics[width=\linewidth]{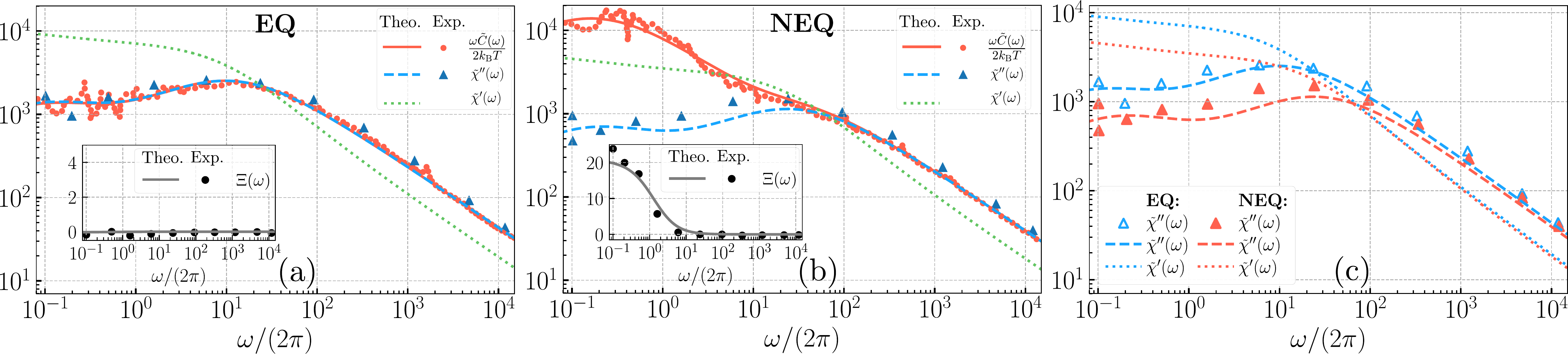}
	\vskip-2mm
\end{figure*}

By summing over the active particles and  introducing  collective variables
$ x(t)={\sum_{i}x_i(t)}/{n}$  and $f(t)={\sum_{i}f_i(t)}/{n}$, we obtain two coupled GLEs  
\begin{subequations}\label{finalGLE}
	\begin{align}
%		\dot{X}(t)&=V(t),\\		
		M\ddot{X}(t)=&-\gamma\int_{-\infty}^{t}{\mathrm{d}}t'\, \mathcal{K}_+(t-t')\dot{X}(t') -(nk+K)X(t)\nonumber\\&+nk x(t)+F(t),\\
%		\dot{x}(t)&=v(t),\\		
		m\ddot{x}(t)=&-\gamma_\text{a}\int_{-\infty}^{t}{\mathrm{d}}t'\, \mathcal{K}_+(t-t')\dot{x}(t') -k(x(t)-X(t))\nonumber\\&+ f(t), 
	\end{align}
\end{subequations}
for the tracer and the collective active-particle coordinate. 
The collective random noise has zero mean  and 
strength  $\langle f(t)f(t')\rangle =  k_\text{B}T_\text{a}\; \gamma_\text{a}\mathcal{K}_+(|t-t'|)/n$.

By Fourier transforming, Eq.\eqref{finalGLE} can be solved for $\tilde{X}(\omega)=\int\! \mathrm{d}t\, X(t)\,{\text e}^{\icomplex\omega t}$ to yield
\begin{eqnarray}
	\tilde{X}(\omega)=\frac{\tilde{F}_\text{R}(\omega)}{-M\omega^2+\icomplex\omega\tilde{\mathcal{K}}_\text{tot}(\omega)+K},\label{totGLE}
\end{eqnarray}
with the effective friction kernel and the effective  noise given by
\begin{eqnarray}
&&	\tilde{\mathcal{K}}_\text{tot}(\omega) = \bigg[\gamma \tilde{\mathcal{K}}_+(\omega)+\frac{nk\gamma_{\text{a}}\tilde{\mathcal{K}}_+(\omega)}{\icomplex\omega\gamma_{\text{a}}\tilde{\mathcal{K}}_+(\omega)+k}\bigg],\label{totMemory} \\
&&	\tilde F_\text{R}(\omega)=\frac{n k \tilde{f}(\omega)}{\icomplex\omega\gamma_{\text{a}}\tilde{\mathcal{K}}_+(\omega)+ k}+ \tilde{F}(\omega). \label{totForce}
\end{eqnarray}

From Eq.~\eqref{totGLE} we can  directly calculate the  positional autocorrelation function, 
$\tilde C(\omega)=\int\! \mathrm{d}t\,\langle X(0)X(t)\rangle \, {\text e}^{\icomplex\omega t}$. 
From the  linear-response   relation
$X(t)=\int_{0}^{\infty}\mathrm{d}t'\chi(t-t')F_\text{R}(t')$,
 the imaginary part of the response function $\tilde\tilde{\chi}''(\omega)$ follows,  see Supplemental Material \cite{SM} for explicit
 derivations. 
 The FDT violation can be quantified by the spectral function \cite{Netz2018}
\begin{equation}
	\Xi(\omega)=\frac{\omega\tilde{C}(\omega)/(2k_{\text{B}}T)}{\tilde{\chi}''(\omega)}-1=\frac{\tilde{C}_F(\omega)/(2k_{\text{B}}T)}{\tilde{\mathcal{K}}'_\text{tot}(\omega)}-1,\label{spectral} 
\end{equation}
where $\tilde{C}_F(\omega) = \int\! \mathrm{d}t\,\langle F_\text{R}(0)F_\text{R}(t)\rangle \,{\text e}^{\icomplex\omega t}$  
is the spectral noise autocorrelation and $\tilde{\mathcal{K}}'_\text{tot}(\omega)$  the real part of the Fourier-transformed memory kernel \cite{SM}. For an equilibrium system characterized by  $\alpha=0$, 
i.e. when the  active and tracer particles  have the same temperature, the FDT is recovered  \cite{Kubo2012} and  $\Xi(\omega)=0$. 
 Assuming overdamped motion with $M\rightarrow0$ and $m\rightarrow0$,  we obtain explicitly
\begin{widetext}
	\begin{eqnarray}
		\label{eq:powerlawchi}
		\tilde{\chi}''(\omega)=&&\,\omega ^{\kappa } \sin \left(\frac{\kappa\pi }{2}\right) \bigg[(\gamma +\hat{\gamma}_\text{a})\hat{k}^2+2 \gamma \hat{\gamma}_\text{a}  \hat{k}\cos\left(\frac{\kappa\pi }{2}\right)\omega ^{\kappa } +\gamma\hat{\gamma}_\text{a}^2  \omega ^{2 \kappa }\bigg]\bigg\{
		\hat{k}^2 K^2+ 	\gamma ^2 \hat{\gamma}_\text{a}^2 \omega ^{4 \kappa }
		\\&&
		+\big[2 \gamma \hat{\gamma}_\text{a} \hat{k}K \cos (\kappa\pi )+\left(\hat{\gamma}_\text{a} (\hat{k}+K)+\gamma  \hat{k}\right)^2\big] \omega ^{2 \kappa }+2 \cos \left(\frac{\kappa\pi }{2}\right) \big[\hat{\gamma}_\text{a} (\hat{k}+K)+\gamma  \hat{k}\big] \big[\hat{k}K+\gamma  \hat{\gamma}_\text{a} \omega ^{2 \kappa }\big] \omega ^{\kappa }\bigg\}^{-1}, \nonumber\\
		\label{eq:powerlawcorr} 
		\tilde C(\omega)=&&\,2k_{\text B}T\omega ^{\kappa-1} \sin \left(\frac{\kappa\pi }{2}\right) \bigg[(\gamma +\hat{\gamma}_\text{a}(1+\alpha))\hat{k}^2+2 \gamma \hat{\gamma}_\text{a}  \hat{k}\cos\left(\frac{\kappa\pi }{2}\right)\omega ^{\kappa } +\gamma\hat{\gamma}_\text{a}^2  \omega ^{2 \kappa }\bigg]\bigg\{
		\hat{k}^2 K^2+ 	\gamma ^2 \hat{\gamma}_\text{a}^2 \omega ^{4 \kappa }
		\\&&
		+\big[2 \gamma \hat{\gamma}_\text{a} \hat{k}K \cos (\kappa\pi )+\left(\hat{\gamma}
		_\text{a} (\hat{k}+K)+\gamma  \hat{k}\right){}^2\big] \omega ^{2 \kappa }+2 \cos \left(\frac{\kappa\pi }{2}\right) \big[\hat{\gamma}_\text{a} (\hat{k}+K)+\gamma  \hat{k}\big] \big[\hat{k}K+\gamma  \hat{\gamma}_\text{a} \omega ^{2 \kappa }\big] \omega ^{\kappa }\bigg\}^{-1},
		\nonumber
	\end{eqnarray}
	\vskip-3mm
	\begin{equation}
		\Xi(\omega)=\frac{\hat{\gamma}_\text{a} \hat{k}^2\alpha}{(\gamma
			+\hat{\gamma}_\text{a})\hat{k}^2 +2 \gamma\hat{\gamma}_\text{a} \hat{k}\cos \left(\frac{\kappa\pi }{2}\right)\omega ^{\kappa }+\gamma \hat{\gamma}_\text{a}^2 \omega ^{2 \kappa }},  \label{eq:powerlawspec}
	\end{equation}
\end{widetext}
where the active-particle number  $n$ only appears implicitly via $\hat\gamma_{\text{a}}=n\gamma_{\text{a}}$ and $\hat k= nk$, 
so we are left with six  parameters, the non-equilibrium parameter $\alpha$, 
the powerlaw exponent $\kappa$, the tracer and active-particle drag coefficients
$\gamma$ and  $\hat\gamma_{\text{a}}$, the tracer confinement strength $K$ and the tracer-active-particle elastic coupling
strength $\hat k$.
We consider  two recent  microrheology experiments on RBC flickering  \cite{Turlier2016} and cross-linked F-actin networks with myosin II molecular motors \cite{Mizuno2007nonequilibrium}, both reporting FDT violation by independently measuring $\tilde{\chi}''(\omega)$ and $\tilde{C}(\omega)$. 
For both experimental systems, it was shown that the FDT remains valid at short times  or high frequencies, 
while strong FDT deviations are observed at long times or low frequencies. 
We digitize  the experimental data reported for $\tilde{\chi}''(\omega)$ and $\omega \tilde{C}(\omega)/(2k_{\text{B}}T)$ 
and first fit their high-frequency limits, which from Eq. (\ref{eq:powerlawchi}) and Eq. (\ref{eq:powerlawcorr}) are predicted as
$\tilde{\chi}''(\omega) \simeq \sin(\frac{\kappa\pi}{2})/(\gamma\,\omega^{\kappa})$ and $\omega\tilde{C}(\omega)/(2k_\text{B}T) \simeq \sin(\frac{\kappa\pi}{2})/(\gamma\,\omega^{\kappa})$
and obtain  $\kappa$ and $\gamma$ for  each data set. 
Then, we simultaneous fit Eqs.~\eqref{eq:powerlawchi}-\eqref{eq:powerlawcorr}   in the entire frequency range to find the remaining four parameters, the fitting parameters are shown in Table~\ref{table:fitparams}, see Supplemental Material \cite{SM} for details on the fitting procedure and a discussion of the uniqueness of the resulting fit parameters. 

We start with the discussion  of the RBC data, 
where the flickering dynamics of the RBC membrane was investigated by tracking the motion of a tracer bead attached to the RBC membrane in an  equatorial position. In the experiments both the {\em in vivo}  (NEQ) and the ATP-depleted  scenarios were considered \cite{Turlier2016},
we denote the latter scenario as EQ although a finite residual ATP concentration is present, the consequences of which will be discussed further below.
In Fig.~\ref{fig:RBC}a,b  we find excellent agreement between the experimental data and the fits
for  $\tilde{\chi}''(\omega)$ and $\omega \tilde{C}(\omega)/(2k_{\text{B}}T)$ 
for both   EQ and  NEQ scenarios in the entire frequency range.
We also show  the real (storage) part of the response function  $\tilde{\chi}'(\omega)$ from our model as
dotted  lines in Fig.~\ref{fig:RBC}a,b,c,  see Supplemental Material \cite{SM} for the explicit expression. 
 In the EQ case  $\tilde{\chi}''(\omega)$ and $\omega \tilde{C}(\omega)/(2k_{\text{B}}T)$  agree rather nicely 
and display  a weak minimum around $\omega/(2 \pi)=2$ s$^{-1}$, that is caused by the competing
elastic and friction effects,
 see Supplemental Material \cite{SM} for an asymptotic analysis of the response  for low and high frequencies.
 In contrast, in the NEQ case the scaled autocorrelation 
$\omega \tilde{C}(\omega)/(2k_{\text{B}}T)$ is much larger than the response function $\tilde{\chi}''(\omega)$,
indicating strong FDT violation. The spectral function $\Xi(\omega)$ in the insets  illustrates that  
FDT violation occurs predominantly at low frequencies. The comparison 
 in Fig.~\ref{fig:RBC}c demonstrates that both real and imaginary parts of the response function 
increase for NEQ  at low frequencies, i.e. the effective  stiffness of the network
decreases at NEQ  (see Supplemental Material \cite{SM}  for a discussion of the low-/high-frequency asymptotic response). 
The viscoelastic power-law exponents for NEQ and EQ RBC experiments  turn out to be $\kappa=0.936$ and 
$\kappa=0.926$ (see Table~\ref{table:fitparams}), respectively, which fall in the  previously reported range $\kappa=0.92-1.07$ for fresh and aged RBCs \cite{Costa2008}. 
The fitted trap stiffness in the NEQ case, $K=1.66\pm0.03$ pN/$\mu$m, agrees with  the experimentally reported value $1.56\pm0.3$ pN/$\mu$m \cite{Turlier2016},
for the EQ scenario our fitted stiffness is substantially higher, which presumably 
reflects  added stiffness due to the viscoelastic EQ  
environment of the tracer particle. Our fit results for the coupling strength between tracer and active particles $\hat{k}$ is of the same 
order of $K$.
 Also, the fitted values of  the tracer bead friction coefficient  $\gamma$ in Table~\ref{table:fitparams} are close to the experimentally reported value
for the bare tracer bead  $\gamma=0.085$ pN s/$\mu$m\cite{Turlier2016} for both NEQ and EQ scenarios,  the fit results for the active-particle
friction coefficients $\hat{\gamma}_\text{a}$ are of the same order.
The good agreement between the fit parameters  and the known experimental parameters shows that the model and the fitting procedure are  robust.
The non-equilibrium parameter in the NEQ case turns out to be  $\alpha=36$ and thus indicates strong departures from EQ behavior.
In the EQ case we obtain a non-zero value $\alpha\simeq0.985$, which indicates weak NEQ due to
 incomplete ATP depletion in the experiment,
  as  reflected by weak deviation between  $\tilde{\chi}''(\omega)$ and $\omega \tilde{C}(\omega)/(2k_{\text{B}}T)$ 
 in Fig.~\ref{fig:RBC}a.

Mizuno et al. \cite{Mizuno2007nonequilibrium} have studied the  {\em in vitro}  mechanics of active cross-linked actomyosin networks in the presence and absence of ATP, again  referred to as NEQ and EQ cases.

The myosin-II motor proteins  
perform stepwise motion along the actin filaments  in the presence of ATP, which
causes sliding motion between two filaments connected by motors and thereby produces contractile active forces in the network.  
As shown in Fig.~\ref{fig:ACM}a, b,  we again find excellent agreement between the experimental data and our fits
for  $\tilde{\chi}''(\omega)$ and $\omega \tilde{C}(\omega)/(2k_{\text{B}}T)$ 
for both the  EQ and the NEQ scenarios in the entire frequency range.
 In the EQ case  $\tilde{\chi}''(\omega)$ and $\omega \tilde{C}(\omega)/(2k_{\text{B}}T)$  are practically identical 
 and show (similar to RBC)  two weak maxima that are well produced by the model,
  see Supplemental Material \cite{SM} for an asymptotic analysis.
In the NEQ case the scaled autocorrelation 
$\omega \tilde{C}(\omega)/(2k_{\text{B}}T)$ is much larger than the response function $\tilde{\chi}''(\omega)$,
indicating strong FDT violation with a magnitude rather similar to the RBC case. 
For the EQ case the  value $\alpha = -0.035$ indicates equilibrium in the absence of ATP. 
In both EQ and NEQ cases the fitted power-law exponent in Table~\ref{table:fitparams} is in excellent agreement with 
$\kappa=3/4$ predicted by  theory \cite{Morse1998Viscoelasticity,Schnurr1997Determining,Hiraiwa2018systematic} and experiments \cite{Mizuno2007nonequilibrium}. 
The comparison 
 in Fig.~\ref{fig:RBC}c demonstrates that both real and imaginary parts of the response function 
 are at low frequencies for the NEQ case smaller than for the EQ case, i.e. the effective  stiffness of the network
increases in the NEQ case, in contrast to the RBC case. 
This behavior is primarily caused by the elastic model parameters:
 we see that $K$, which describes the stiffness of the tracer bead confinement, 
 and $\hat{k}$, which describes the coupling strength between the tracer and the active particles,
 increase for actomyosin networks when going from the EQ to the NEQ scenario, whereas the opposite trend is observed
 for  the RBC flickering data, see Table~\ref{table:fitparams}.

In summary,
we introduce a simple model for viscoelastic active biomatter  that incorporates the elastic coupling between a tracer bead and active particles and 
the  medium viscoelasticity through a memory kernel and yields closed-form analytical predictions for the autocorrelation and response function
of the  tracer bead. 
These predictions are in excellent agreement with  experimental data for RBCs and actomyosin networks and in particular allow
to quantify the departure from EQ   by the NEQ parameter $\alpha$, which corresponds to the 
effective temperature  ratio of the reservoirs coupled to the active particles and the tracer bead in our model. We find $\alpha \approx 23$ and 36 
in the presence of ATP and $\alpha \approx 0$ and 1 in the absence of ATP in  the two experimental systems
we considered, clearly signaling the strong NEQ character induced by ATP.
Note that the effective temperature of the active particles does not correspond to the actual physical temperature, 
but rather characterizes the strength of the stochastic force the motor proteins produce and the power they dissipate.
Our model has a wide range of applicability for active biomaterials and experimental setups and  with suitable
modifications can for example predict cross-correlations in two-particle microrheology experiments \cite{Mizuno2009high}.

We acknowledge H{\'e}l{\`e}ne Berthoumieux for useful comments on manuscript. A.A. and R.R.N.  acknowledge funding from the European Research Council (ERC) under the European Union's Horizon 2020 Research and Innovation Program under Grant Agreement No. 835117.  A.A. acknowledges the
hospitality by the  Institute for Research in Fundamental Sciences (IPM), Tehran, during  the initial stages of this work.
\bibliography{biblio}% Produces the bibliography via BibTeX.

\end{document}

\title{{\em Supplemental Material for the article}:\\Non-Markovian modeling of non-equilibrium fluctuations and dissipation in active viscoelastic biomatter}
	
	\author{Amir Abbasi}
	\affiliation{
		Fachbereich Physik, Freie Universit\"at Berlin, Arnimallee 14, 14195 Berlin, Germany}
	\author{Roland R. Netz}
	\affiliation{
		Fachbereich Physik, Freie Universit\"at Berlin, Arnimallee 14, 14195 Berlin, Germany}%
	
	\author{Ali Naji}
	\thanks{a.naji@ipm.ir}
\affiliation{School of Nano Science, Institute for Research in Fundamental Sciences (IPM), Tehran 19395-5531, Iran}
	
%	\date{\today}
	\maketitle
	
	%%%%%%%%%%%%%%%%%%%%%%%%%%%%%%%%%%%%%%%%%%%%%%%%%%%%%%%%%%%%%%%%%%%%%%%%%%%%%%%%%%%%%%%%%%%%%%%%%%%%%%%%%%%%%%%%%%%%%%%%%%%%%%%%%%%%%%%%%%%%%%%%%%%%%%%%%%%%%%%%%%%%%%%%%%%%%%%%%
	\section{\label{supp:spectral}Spectral function derivation}
	Here we derive Eq.~(7) in the main text. By Fourier transforming the set of coupled GLEs introduced in Eq.~(3) in the main text, we have
	\begin{subequations}\label{eq:GLEsysfourier}
		\begin{align}		
			-M\omega^2\tilde{X}(\omega)&=\icomplex\omega \gamma \tilde{\mathcal{K}}_+(\omega)\tilde{X}(\omega) -(nk+K)\tilde{X}(\omega)+nk\tilde{x}(\omega)+ \tilde{F}(\omega)+\tilde{F}_\text{ext}(\omega),\\		
			-m\omega^2\tilde{x}(\omega)&=\icomplex\omega\gamma_\text{a} \tilde{\mathcal{K}}_+(\omega)\tilde{x}(\omega) -k(\tilde{x}(\omega)-\tilde{X}(\omega))+\tilde{f}(\omega).
		\end{align}
	\end{subequations}
	A tilde indicates a Fourier-transformed quantity and we use $\alpha$ to quantify departure from equilibrium.
	Solving the second equation for  $\tilde{x}(\omega)$ in the over-damped limit, i.e., for $m\rightarrow0$, we arrive at
	\begin{equation}
		\tilde{x}(\omega)=\frac{k X(\omega )+\tilde{f}(\omega )}{k-\icomplex \omega\gamma_\text{a}\tilde{\mathcal{K}}_+(\omega)}.\label{eq:activepos}
	\end{equation}
	Substituting the expression above for $\tilde x(\omega)$ in Eq.~\eqref{eq:GLEsysfourier}(a) and after some intermediate steps we arrive at
	\begin{equation}
		-M\omega^2\tilde{X}(\omega)=-K\tilde{X}(\omega)+\icomplex\omega\tilde{\mathcal{K}}_\text{tot}(\omega)\tilde{X}(\omega)+\tilde{F}_\text{R}(\omega)+\tilde{F}_\text{ext}(\omega),\label{eq:X_omega}
	\end{equation}
	where
	\begin{equation}
		\tilde{\mathcal{K}}_\text{tot}(\omega)=\gamma \tilde{\mathcal{K}}_+(\omega)+\frac{nk\gamma_\text{a}\tilde{\mathcal{K}}_+(\omega)}{k-\icomplex \omega\gamma_\text{a}\tilde{\mathcal{K}}_+(\omega)},\label{eq:total_kernel}
	\end{equation}
	and
	\begin{equation}
		\tilde{F}_\text{R}(\omega)=\frac{nk\tilde f(\omega )}{k-\icomplex \omega\gamma_\text{a}\tilde{\mathcal{K}}_+(\omega)}+ \tilde{F}(\omega).
	\end{equation}
	The response function in the time domain  is defined by the relation
	\begin{equation}
		X(t)=\int_{-\infty}^{+\infty}{\mathrm{d}}t'\chi(t-t')F_{\text{ext}}(t'),\label{eq:def_response}
	\end{equation}  
	where due to causality, we have $\chi(t)=0$ for $t<0$. Taking the Fourier transform and averaging over the random noise we have
	\begin{equation}
		\tilde{\chi}(\omega)=\frac{\langle\tilde{X}(\omega)\rangle}{\tilde F_{\text{ext}}(\omega)}.\label{eq:chidef}
	\end{equation}
	Using this definition  for the response function and  Eq.~\eqref{eq:X_omega}, we arrive at  
	\begin{equation}
		\tilde{\chi}(\omega) = \bigg[-M\omega^2+K-\icomplex\omega\tilde{\mathcal{K}}_\text{tot}(\omega)\bigg]^{-1}.  \label{eq:response}
	\end{equation}
	Now we derive the total random force correlation function, which in the Fourier domain reads
	\begin{equation}
		\langle\tilde{F}_\text{R}(\omega)\tilde{F}_\text{R}(\omega')\rangle=\bigg[\frac{n^2k^2\langle\tilde f(\omega)f(\omega')\rangle}{[k-\icomplex \omega\gamma_\text{a}\tilde{\mathcal{K}}_+(\omega)][k-\icomplex \omega'\gamma_\text{a}\tilde{\mathcal{K}}_+(\omega')]}+\langle\tilde{F}(\omega)\tilde{F}(\omega')\rangle \bigg],\label{eq:F_RR}
	\end{equation}
	where
	\begin{eqnarray}
&&		\langle\tilde{F}(\omega)\tilde{F}(\omega')\rangle =\int_{-\infty}^{\infty}\mathrm{d}t'\int_{-\infty}^{\infty}\mathrm{d}t \,\langle F(t)F(t')\rangle e^{\icomplex \omega t} e^{\icomplex \omega' t'}\nonumber\\
&&		= k_\text{B}T\int_{-\infty}^{\infty}\mathrm{d}t'\int_{-\infty}^{\infty}\mathrm{d}t\, \mathcal{K}_+(|t-t'|)\, e^{\icomplex \omega t} e^{\icomplex \omega' t'}=k_\text{B}T\int_{-\infty}^{\infty}\mathrm{d}t' e^{\icomplex t' (\omega+\omega')}\int_{-\infty}^{\infty}\mathrm{d}u\, \mathcal{K}_+(|u|)\, e^{\icomplex \omega u}\nonumber\\
&&		= 2\pi k_\text{B}T\delta(\omega+\omega')\big[\tilde{\mathcal{K}}_+(\omega)+\tilde{\mathcal{K}}_+(-\omega)\big]=2\pi k_\text{B}T\delta(\omega+\omega')\tilde{\mathcal{K}}(\omega).\label{eq:F_corr}
	\end{eqnarray}
	and
	\begin{eqnarray}
&&		\langle\tilde{f}(\omega)\tilde{f}(\omega')\rangle=\int_{-\infty}^{\infty}\mathrm{d}t'\int_{-\infty}^{\infty}\mathrm{d}t \,\langle f(t)f(t')\rangle e^{\icomplex \omega t} e^{\icomplex \omega' t'}\nonumber\\
&&		=\int_{-\infty}^{\infty}\mathrm{d}t'\int_{-\infty}^{\infty}\mathrm{d}t\big\langle \frac{\sum_{i=1}^nf_i(t)}{n}\frac{\sum_{j=1}^nf_j(t')}{n}\big\rangle=\int_{-\infty}^{\infty}\mathrm{d}t'\int_{-\infty}^{\infty}\mathrm{d}t \frac{\sum_{i,j=1}^n\langle f_i(t)f_j(t')\rangle}{n^2}\nonumber\\
&&		=\int_{-\infty}^{\infty}\mathrm{d}t'\int_{-\infty}^{\infty}\mathrm{d}t \frac{\sum_{i,j=1}^nk_\text{B}T_\text{a}\gamma_\text{a}\delta_{ij}\mathcal{K}_+(|t-t'|)}{n^2}=\frac{k_\text{B}T_\text{a}\gamma_{\text{a}}}{n}\int_{-\infty}^{\infty}\mathrm{d}t'\int_{-\infty}^{\infty}\mathrm{d}t\, \mathcal{K}_+(|t-t'|)\, e^{\icomplex \omega t} e^{\icomplex \omega' t'}\nonumber\\
&&		=\frac{k_\text{B}T_\text{a}\gamma_{\text{a}}}{n}\int_{-\infty}^{\infty}\mathrm{d}t' e^{\icomplex t' (\omega+\omega')}\int_{-\infty}^{\infty}\mathrm{d}u\, \mathcal{K}_+(|u|)\, e^{\icomplex \omega u}=2\pi\frac{k_\text{B}T_\text{a}\gamma_{\text{a}}}{n}\delta(\omega+\omega')\tilde{\mathcal{K}}(\omega)\nonumber
	\end{eqnarray}
	In the last step we changed variables as $u=t-t'$ and  defined $\tilde{\mathcal{K}}(\omega)\equiv\tilde{\mathcal{K}}_+(\omega)+\tilde{\mathcal{K}}_+(-\omega)=2\tilde{\mathcal{K}}'_+(\omega)$. Thus, in the  time domain we have
	\begin{eqnarray}\label{eq:tot_force_corr}
&&		C_F(t)=\langle F_\text{R}(t_0)F_\text{R}(t_0+t)\rangle =\int_{-\infty}^{\infty}\frac{\mathrm{d}\omega}{2\pi}\int_{-\infty}^{\infty}\frac{\mathrm{d}\omega'}{2\pi}\langle\tilde{F}_\text{R}(\omega)\tilde{F}_\text{R}(\omega')\rangle e^{-\icomplex\omega' t_0}e^{-\icomplex\omega (t_0+t)}\nonumber\\
&&		= \int_{-\infty}^{\infty}\frac{\mathrm{d}\omega}{2\pi}\int_{-\infty}^{\infty}\frac{\mathrm{d}\omega'}{2\pi}\bigg[\frac{nk^2\,\gamma_{\text{a}}k_\text{B}T_\text{a}}{[k-\icomplex \omega\gamma_\text{a}\tilde{\mathcal{K}}_+(\omega)][k-\icomplex \omega'\gamma_\text{a}\tilde{\mathcal{K}}_+(\omega')]}+k_\text{B}T\bigg] 2\pi\delta(\omega+\omega')\tilde{\mathcal{K}}(\omega)e^{-\icomplex\omega' t_0}e^{-\icomplex\omega (t_0+t)}\nonumber\\
&&		= \int_{-\infty}^{\infty}\frac{\mathrm{d}\omega}{2\pi}\underbrace{\bigg[\frac{nk^2\,\gamma_{\text{a}}k_\text{B}T_\text{a}}{[k-\icomplex \omega\gamma_\text{a}\tilde{\mathcal{K}}_+(\omega)][k-\icomplex \omega'\gamma_\text{a}\tilde{\mathcal{K}}_+(\omega')]}+k_\text{B}T\bigg] \tilde{\mathcal{K}}(\omega)}_{\tilde{C}_F(\omega)} e^{-\icomplex\omega t}.
	\end{eqnarray}
	In the same fashion, using Eqs.~\ref{eq:X_omega}, \ref{eq:response} and setting $\tilde{F}_\text{ext}(\omega)=0$, we define
	\begin{eqnarray}
		C(t)&=&\langle X(t_0)X(t_0+t)\rangle =\int_{-\infty}^{\infty}\frac{\mathrm{d}\omega}{2\pi}\int_{-\infty}^{\infty}\frac{\mathrm{d}\omega'}{2\pi}\langle\tilde{X}(\omega)\tilde{X}(\omega')\rangle e^{-\icomplex\omega' t_0}e^{-\icomplex\omega (t_0+t)}\\
		&=&\int_{-\infty}^{\infty}\frac{\mathrm{d}\omega}{2\pi}\int_{-\infty}^{\infty}\frac{\mathrm{d}\omega'}{2\pi}\tilde{\chi}(\omega)\tilde{\chi}(\omega')\langle \tilde{F}_\text{R}(\omega)\tilde{F}_\text{R}(\omega')\rangle  e^{-\icomplex\omega' t_0}e^{-\icomplex\omega (t_0+t)}\nonumber\\
		&=&\int_{-\infty}^{\infty}\frac{\mathrm{d}\omega}{2\pi}\int_{-\infty}^{\infty}\frac{\mathrm{d}\omega'}{2\pi}\tilde{\chi}(\omega)\tilde{\chi}(\omega')\,\tilde{C}_F(\omega)\,\delta(\omega+\omega')  e^{-\icomplex\omega' t_0}e^{-\icomplex\omega (t_0+t)}\nonumber\\
		&=&\int_{-\infty}^{\infty}\frac{\mathrm{d}\omega}{2\pi}\underbrace{\tilde{\chi}(\omega)\tilde{\chi}(-\omega)\,\tilde{C}_F(\omega)}_{\tilde{C}(\omega)} 
		e^{-\icomplex\omega t}.\label{eq:poscorr}
	\end{eqnarray}
	We can split the response function given in Eq.~\eqref{eq:response} into its real and imaginary parts as
	\begin{equation}
	\tilde{\chi}'(\omega)+\icomplex\tilde{\chi}''(\omega) =\frac{1}{A-\icomplex\omega\tilde{\mathcal{K}}_\text{tot}'(\omega)}\frac{A+\icomplex\omega\tilde{\mathcal{K}}_\text{tot}'(\omega)}{A+\icomplex\omega\tilde{\mathcal{K}}_\text{tot}'(\omega)},
	\end{equation}
	where $A$ is the real part of the denominator given by
	\begin{equation}
		A=-M\omega^2+K+\omega\tilde{\mathcal{K}}_\text{tot}''(\omega).
	\end{equation}
	Now, for the dissipative part of the response function, we have
	\begin{equation}
		\tilde{\chi}''(\omega)=\frac{\omega\tilde{\mathcal{K}}_\text{tot}'(\omega)}{\big[A-\icomplex\omega\tilde{\mathcal{K}}_\text{tot}'(\omega)\big]\big[A+\icomplex\omega\tilde{\mathcal{K}}_\text{tot}'(\omega)\big]}=\omega\tilde{\mathcal{K}}_\text{tot}'(\omega)\tilde{\chi}(\omega)\tilde{\chi}^*(\omega).\label{eq:chipp}
	\end{equation}
	The Fourier transform of response function reads
	\begin{equation}
		\tilde{\chi}(\omega)=\int_{-\infty}^{\infty}\mathrm{d}t\chi(t)e^{\icomplex\omega t}.\label{eq:chi_omega}
	\end{equation}
	where $\chi(t)=0$ for $t<0$ due to causality. Since $\chi(t)$ is a real function one has $\tilde{\chi}^*(\omega)=\tilde{\chi}(-\omega)$. 
	Thus, Eq.~\eqref{eq:chipp} yields
	\begin{equation}
		\tilde{\chi}''(\omega) =\omega\tilde{\mathcal{K}}_\text{tot}'(\omega)\tilde{\chi}(\omega)\tilde{\chi}(-\omega),\label{eq:kernelreal}
	\end{equation}
	and from Eq.~\eqref{eq:poscorr} we have
	\begin{equation}
		\tilde{C}(\omega) = \tilde{\chi}(\omega)\tilde{\chi}(-\omega)\tilde{C}_F(\omega).\label{eq:poscorrlast}
	\end{equation}
	Combining Eqs.~\ref{eq:kernelreal} and \ref{eq:poscorrlast} we arrive at 
	\begin{equation}
		\Xi(\omega)=\frac{\omega\tilde{C}(\omega)/(2k_{\text{B}}T)}{\tilde{\chi}''(\omega)}-1=\frac{\tilde{C}_F(\omega)/(2k_{\text{B}}T)}{\tilde{\mathcal{K}}'_\text{tot}(\omega)}-1.\label{eq:spectral}
	\end{equation}
	Using the fluctuation-dissipation theorem in the time domain, 
	$\chi(t)=-\theta(t)\dot C(t)/(k_\text{B}T)$, where $ \theta(t)$ denotes the Heavyside function, in the Fourier domain we have
	\begin{equation}
		\tilde{\chi}''(\omega)=\frac{\omega\tilde{C}(\omega)}{2k_\text{B}T}.\label{eq:equilfdt}
	\end{equation}
	Comparing Eqs.~\ref{eq:spectral} and \ref{eq:equilfdt} we realize that in equilibrium we have $\Xi(\omega)=0$.
	%%%%%%%%%%%%%%%%%%%%%%%%%%%%%%%%%%%%%%%%%%%%%%%%%%%%%%%%%%%%%%%%%%%%%%%%%%%%%%%%%%%%%%%%%%%%%%%%%%%%%%%%%%%%%%%%%%%%%%%%%%%%%%%%%%%%%%%%%%%%%%%%%%%%%%%%%%%%%%%%%%%%%%%%%%%%%%%%%%%%%%%%%%%%%%%%%%%%%%%%%%%%%%%%%%%%%%%%%%%%%%%%%%%%%%%%%%%%%%%%%%%%%%%%%%%%%%%%%%%%%%%%%%%%%%%%%%%%%%%%%%%%%%%%%%%%%%%%%%%%%%%
	In the next step, we show that departure from equilibrium happens when $\alpha$ deviates from zero. 
	Using Eqs.~\ref{eq:total_kernel}, \ref{eq:tot_force_corr} and \ref{eq:spectral}, after some intermediate steps we find
	\begin{eqnarray}
		\Xi(\omega)&=&\frac{\tilde{C}_F(\omega)/(2k_{\text{B}}T)}{\tilde{\mathcal{K}}'_\text{tot}(\omega)}-1=\frac{k^2\gamma_{\text{a}}\alpha}{k^2(\gamma+\gamma_{\text{a}})-2k\gamma\gamma_{\text{a}}\omega\,\mathcal{K}_+''(\omega)+\gamma\gamma_{\text{a}}^2\,\omega^2\,|\mathcal{K}_+(\omega)|^2},
	\end{eqnarray}
	where we have used the definition
	$T_{\text{a}}\equiv T(\alpha+1)$. Thus to recover the equilibrium state, i.e. $\Xi(\omega)=0$, the NEQ parameter $\alpha$ should be equal to zero. 
	
	%%%%%%%%%%%%%%%%%%%%%%%%%%%%%%%%%%%%%%%%%%%%%%%%%%%%%%%%%%%%%%%%%%%%%%%%%%%%%%%%%%%%%%%%%%%%%%%%%%%%%%%%%%%%%%%%%%%%%%%%%%%%%%%%%%%%%%%%%%%%%%%%%%%%%%%%%%%%%%%%%%%%%%%%%%%%%%%%%
	
	\section{\label{supp:Allmodels}Response functions, positional autocorrelations and spectral functions for different models}
	Using memory kernels for different viscoelastic models (see Table~\ref{table:kernels}), we can derive different expressions for the response  
	and positional autocorrelation functions. 
	\subsection{\label{subsec:deriv_memory}Derivation of memory kernels in Fourier space}
	
	Here we briefly review the derivation of the Fourier transform of the memory kernels used in this paper.
	
	\textbf{Newtonian fluid}:
	\begin{equation}
		\int_{0}^{+\infty} \mathrm{d}t \, 2 \gamma\delta(t) e^{\icomplex \omega t} = \gamma
	\end{equation}
	
	\textbf{Maxwell model}:
	\begin{equation}
		\int_{0}^{+\infty} \mathrm{d}t \, \frac{\gamma}{\tau_\text{M}} e^{-t/\tau_{\text{M}}} e^{\icomplex \omega t} =\frac{\gamma}{1-\icomplex\omega\tau_\text{M}}
	\end{equation}
	
	\textbf{Kelvin-Voigt model}:
	\begin{equation}
		\int_{0}^{+\infty} \mathrm{d}t \, 2\gamma\delta(t) e^{\icomplex \omega t} = \gamma 
	\end{equation}
	We treat the elastic term of the Kelvin-Voigt model as a potential contribution, which stays outside the memory kernel. 
	Therefore, the memory contribution of the Kelvin-Voigt model is the same as that of the Newtonian fluid.

	\textbf{Power-law model}:
	\begin{equation}
		\int_{0}^{+\infty} \mathrm{d}t \, \frac{\gamma \,t^{-\kappa}}{\Gamma(1-\kappa)} e^{\icomplex \omega t}= \frac{\gamma(-\icomplex\omega)^\kappa}{-\icomplex\omega\Gamma(1-\kappa)}\int_{0}^{+\infty}\zeta^{-\kappa}e^{-\zeta}d\zeta=\gamma(-\icomplex\omega)^{\kappa-1}=\gamma\omega^{\kappa-1}\big[\sin(\frac{\kappa\pi}{2})+\icomplex \cos(\frac{\kappa\pi}{2})\big],  \label{eq:pow_kernel}
	\end{equation}
	where we have changed variables as $\zeta=-\icomplex\omega t$. We note that the integral in Eq.~\eqref{eq:pow_kernel} converges only for $\kappa<1$.
	
	\begin{table}
		\begin{ruledtabular}
			\begin{tabular}{lcc}
				\textrm{Model}&
				\textrm{Memory kernel, $\gamma\mathcal{K}_+(t)$}&
				\textrm{Memory kernel, $\gamma\tilde{\mathcal{K}}_+(\omega)$}\\
				\colrule
				Newtonian & $2\gamma \delta(t)$ & $\gamma$  \\ 
				Maxwell & $(\gamma/\tau_{\text{M}})\, e^{-t/\tau_{\text{M}}}$  & $\gamma/(1-\icomplex\omega\tau_{\text{M}})$ \\
				Kelvin-Voigt & $2\gamma\delta(t)$ & $\gamma$\\
				Power-law & $\gamma \,t^{-\kappa}/\Gamma(1-\kappa)$ & $\gamma(\icomplex\omega)^{\kappa-1}$ \\
			\end{tabular}
		\end{ruledtabular}
		\caption{\label{table:kernels} Memory kernels for different viscoelastic models and their Fourier transform. To obtain the memory kernel for active particles, one has to replace $\gamma\rightarrow\gamma_{\text{a}}$ everywhere in this table. Note that for the Kelvin-Voigt model one also has to  replace $k_\text{KV}\rightarrow k_\text{KV,a}$, 	 see Sec.~\ref{subsec:deriv_memory} for more details.
		}
	\end{table}
	Now, exploiting the memory kernels introduced above, we obtain  expressions for the real part of the response function $\tilde\chi'(\omega)$, the imaginary part of the response function $\tilde\chi''(\omega)$, the positional correlation function $\tilde C(\omega)$ and the spectral function $\Xi(\omega)$.
	
	\subsection{Newtonian fluid}
	\begin{eqnarray}
		\tilde\chi'(\omega)=&&\frac{\hat{k}^2 K+(K+\hat{k})\hat{\gamma}_{\text{a}}^2\omega^2}{\hat{k}^2
			\big[\omega ^2 (\hat{\gamma}_\text{a}+\gamma
			)^2+K^2\big]+2 \hat{k} K \omega ^2 \hat{\gamma}	_\text{a}^2 +\omega ^2 \hat{\gamma} _\text{a}^2 (\gamma ^2
			\omega ^2+K^2)},\label{Newtonchip}
	\end{eqnarray}
	\begin{eqnarray}
		\tilde\chi''(\omega)=&&\frac{\hat{k}^2 (\hat{\gamma} _\text{a}+\gamma)\omega+\gamma\hat{\gamma}_\text{a}^2 \omega ^3}{\hat{k}^2
			\big[\omega ^2 (\hat{\gamma}_\text{a}+\gamma
			)^2+K^2\big]+2 \hat{k} K \omega ^2 \hat{\gamma}
			_\text{a}^2+\omega ^2 \hat{\gamma} _\text{a}^2 (\gamma ^2
			\omega ^2+K^2)},\label{Newtonchi}
	\end{eqnarray}
	\begin{eqnarray}
		\tilde{C}(\omega)=&&\frac{2k_\text{B}T \big[\gamma \hat{\gamma}_\text{a}^2 \omega ^2 +\hat{k}^2[\gamma+\hat{\gamma} _\text{a}(1+\alpha)]\big]}{\hat{k}^2
			\big[\omega ^2 (\hat{\gamma}_\text{a}+\gamma
			)^2+K^2\big]+2 \hat{k} K \omega ^2 \hat{\gamma}
			_\text{a}^2+\omega ^2 \hat{\gamma} _\text{a}^2 (\gamma ^2
			\omega ^2+K^2)},\label{Newtoncorr}
	\end{eqnarray}
	\begin{equation}
		\Xi(\omega)=\frac{\hat{k}^2\hat{\gamma}_\text{a} \alpha}{\hat{k}^2(\hat{\gamma}
			_\text{a}+\gamma)+\gamma \hat{\gamma}_\text{a}^2 \omega ^2}.\label{Newtonspec}
	\end{equation}
	%%%%%%%%%%%%%%%%%%%%%%%%%%%%%%%%%%%%%%%%%%%%%%%%%%%%%%%%%%%%%%%%%%%%%%%%%%%%%%%%%%%%%%%%%%%%%%%%%%%%%%%%%%%%%%%%%%%%%%%%%%%%%%%%%%%%%%%%%%%%%%%%%%%%%%%%%%%%%%%%%%%%%%%%%%%%%%%%%%%%%%%%%%%%%%%%%%%%%%%%%%%%%%%
	\subsection{Maxwell model}
	\begin{eqnarray}
		\tilde\chi'(\omega)=&&\bigg[\hat{k}^2 K+\left(K \hat{\gamma}_\text{a}^2+\hat{k} \hat{\gamma}_\text{a} \left(\hat{\gamma}_\text{a}+2 K \tau_\text{M}\right)+\hat{k}^2 \tau_\text{M} \left(\gamma +\hat{\gamma}_\text{a}+2 K
		\tau_\text{M}\right)\right) \omega ^2+\tau_\text{M} \left(\hat{\gamma}_\text{a}+\hat{k} \tau_\text{M}\right) \nonumber\\
		&&\times\left(\gamma  \left(\hat{\gamma}_\text{a}+\hat{k} \tau
		_M\right)+\tau_\text{M} \left((\hat{k}+K) \hat{\gamma}_\text{a}+\hat{k} K \tau_\text{M}\right)\right) \omega ^4\bigg]\nonumber\\&&\times\bigg\{\hat{k}^2 K^2+ \big[\hat{k}^2 \left(\left(\hat{\gamma}_\text{a}+\gamma \right)^2+2 K \left(\hat{\gamma}
		_\text{a}+\gamma \right) \tau_\text{M}+2 K^2 \tau_\text{M}^2\right)+2 \hat{k} K \hat{\gamma}_\text{a}
		\left(\hat{\gamma}_\text{a}+K \tau_\text{M}\right)+K^2 \hat{\gamma}_\text{a}^2\big]\omega^2 \nonumber\\&&+ \big[\tau_\text{M} \left(\hat{\gamma}_\text{a} (\hat{k}+K)+\hat{k} K \tau_\text{M}\right)+\gamma 
		\left(\hat{\gamma}_\text{a}+\hat{k} \tau_\text{M}\right)\big]^2\omega ^4\bigg\}^{-1},\label{Maxwellchip}
	\end{eqnarray}
	\begin{eqnarray}
		\tilde\chi''(\omega)=&&\bigg[\hat{k}^2 \omega  (\gamma +\hat{\gamma}_\text{a})+
		(k \tau_{\text{M}} +\hat{\gamma}_\text{a})\big[ \hat{k}^2\hat{\gamma}_{\text{a}}\tau_\text{M}^2+\gamma(\hat{\gamma}_\text{a}+\hat{k}\,\tau_\text{M})^2\big]\omega^3\bigg]\nonumber\\&&\times\bigg\{\hat{k}^2 K^2+ \big[\hat{k}^2 \left(\left(\hat{\gamma}_\text{a}+\gamma \right)^2+2 K \left(\hat{\gamma}
		_\text{a}+\gamma \right) \tau_\text{M}+2 K^2 \tau_\text{M}^2\right)+2 \hat{k} K \hat{\gamma}_\text{a}
		\left(\hat{\gamma}_\text{a}+K \tau_\text{M}\right)+K^2 \hat{\gamma}_\text{a}^2\big]\omega^2 \nonumber\\&&+ \big[\tau_\text{M} \left(\hat{\gamma}_\text{a} (\hat{k}+K)+\hat{k} K \tau_\text{M}\right)+\gamma 
		\left(\hat{\gamma}_\text{a}+\hat{k} \tau_\text{M}\right)\big]^2\omega ^4\bigg\}^{-1},\label{Maxwellchi}
	\end{eqnarray}
	\begin{eqnarray}
		\tilde{C}(\omega)=&&2k_\text{B}T\bigg[\hat{k}^2\big[\gamma+\hat{\gamma}_{\text{a}}(1+\alpha)\big]+\big[\hat{k}^2\,\hat{\gamma}_{\text{a}}(1+\alpha)\tau_\text{M}^2+\gamma(\hat{\gamma}_\text{a}+\hat{k}\tau_\text{M})^2\big]\omega^2 \bigg]\nonumber\\&&\times\bigg\{\hat{k}^2 K^2+ \big[\hat{k}^2 \left(\left(\hat{\gamma}_\text{a}+\gamma \right)^2+2 K \left(\hat{\gamma}
		_\text{a}+\gamma \right) \tau_\text{M}+2 K^2 \tau_\text{M}^2\right)+2 \hat{k} K \hat{\gamma}_\text{a}
		\left(\hat{\gamma}_\text{a}+K \tau_\text{M}\right)+K^2 \hat{\gamma}_\text{a}^2\big]\omega^2 \nonumber\\&&+ \big[\tau_\text{M} \left(\hat{\gamma}_\text{a} (\hat{k}+K)+\hat{k} K \tau_\text{M}\right)+\gamma 
		\left(\hat{\gamma}_\text{a}+\hat{k} \tau_\text{M}\right)\big]^2\omega ^4\bigg\}^{-1},\label{Maxwellcorr}
	\end{eqnarray}
	\begin{equation}
		\Xi(\omega)=\frac{\hat{k}^2 \hat{\gamma}_\text{a}\, \alpha \,\left(1+\omega ^2 \tau _\text{M}^2\right)}{\hat{k}^2
			\left(\hat{\gamma}_\text{a}+\gamma \right)+\omega ^2 \left(\hat{k}^2 \hat{\gamma}_\text{a} \tau
			_\text{M}^2+\gamma  \left(\hat{\gamma}_\text{a}+\hat{k} \tau _\text{M}\right)^2\right)}.\label{Maxwellspec}
	\end{equation}
	%%%%%%%%%%%%%%%%%%%%%%%%%%%%%%%%%%%%%%%%%%%%%%%%%%%%%%%%%%%%%%%%%%%%%%%%%%%%%%%%%%%%%%%%%%%%%%%%%%%%%%%%%%%%%%%%%%%%%%%%%%%%%%%%%%%%%%%%%%%%%%%%%%%%%%%%%%%%%%%%%%%%%%%%%%%%%%%%%%%%%%%%%%%%%%%%%%%%%%%%%%%%%%%
	\subsection{Kelvin-Voigt model}	
	\begin{equation}
		\hskip-2cm\tilde{\chi}'(\omega)=\frac{(\hat{k}+\hat{k}_\text{KV,a})\big[\hat{k}_\text{KV,a}(K+k_\text{KV})+\hat{k}(K+k_\text{KV}+\hat{k}_\text{KV,a})\big]+(K+\hat{k}+k_\text{KV})\hat{\gamma}_{\text{a}}^2\omega^2}{\big[(k_{\text{KV}
			}+K) \hat{k}_{\text{KV,a}}+\hat{k}
			(\hat{k}_{\text{KV,a}}+k_{\text{KV}}+K)\big]^2+\big[\gamma ^2
			(\hat{k}_{\text{KV,a}}+\hat{k})^2+2 \gamma \hat{\gamma}_\text{a} \hat{k}^2 +\hat{\gamma}_\text{a}^2
			(k_{\text{KV}}+\hat{k}+K)^2\big]\omega ^2+\gamma ^2
			\hat{\gamma}_\text{a}^2\omega ^4},\label{eq:KVchip}
	\end{equation}
	\begin{equation}
		\hskip-2cm\tilde{\chi}''(\omega)=\frac{  \big[\gamma  (\hat{k}_{\text{KV,a}}^2+\omega ^2
			\hat{\gamma}_\text{a}^2)+2 \gamma  \hat{k} \hat{k}_{\text{KV,a}}+\hat{k}^2 (\hat{\gamma}
			_\text{a}+\gamma )\big]\omega}{\big[(k_{\text{KV}
			}+K) \hat{k}_{\text{KV,a}}+\hat{k}
			(\hat{k}_{\text{KV,a}}+k_{\text{KV}}+K)\big]^2+\big[\gamma ^2
			(\hat{k}_{\text{KV,a}}+\hat{k})^2+2 \gamma \hat{\gamma}_\text{a} \hat{k}^2 +\hat{\gamma}_\text{a}^2
			(k_{\text{KV}}+\hat{k}+K)^2\big]\omega ^2+\gamma ^2
			\hat{\gamma}_\text{a}^2\omega ^4},\label{eq:KVchi}
	\end{equation}
	\begin{eqnarray}
		\hskip-2cm\tilde{C}(\omega)=\frac{2 k_\text{B}T\,\big[\gamma  (\hat{k}_{\text{KV,a}}^2+\omega ^2 \hat{\gamma}
			_\text{a}^2)+2 \gamma \hat{k} \,\hat{k}_{\text{KV,a}}+\hat{k}^2 (\hat{\gamma}
			_\text{a}(1+\alpha)+\gamma )\big]}{\big[(k_{\text{KV}
			}+K) \hat{k}_{\text{KV,a}}+\hat{k}
			(\hat{k}_{\text{KV,a}}+k_{\text{KV}}+K)\big]^2+\big[\gamma ^2
			(\hat{k}_{\text{KV,a}}+\hat{k})^2+2 \gamma \hat{\gamma}_\text{a} \hat{k}^2 +\hat{\gamma}_\text{a}^2
			(k_{\text{KV}}+\hat{k}+K)^2\big]\omega ^2+\gamma ^2
			\hat{\gamma}_\text{a}^2\omega ^4},\nonumber\\\label{KVcorr}
	\end{eqnarray}
	\begin{equation}
		\Xi(\omega)=\frac{\hat{k}^2 \,\hat{\gamma}_\text{a} \alpha}{2 \gamma  \hat{k}
			\hat{k}_{\text{KV,a}}+\hat{k}^2 (\hat{\gamma}_\text{a}+\gamma )+\gamma 
			(\hat{k}_{\text{KV,a}}^2+\hat{\gamma}_\text{a}^2\omega^2)}.\label{KVspec}
	\end{equation}

	%%%%%%%%%%%%%%%%%%%%%%%%%%%%%%%%%%%%%%%%%%%%%%%%%%%%%%%%%%%%%%%%%%%%%%%%%%%%%%%%%%%%%%%%%%%%%%%%%%%%%%%%%%%%%%%%%%%%%%%%%%%%%%%%%%%%%%%%%%%%%%%%%%%%%%%%%%%%%%%%%%%%%%%%%%%%%%%%%%%%%%%%%%%%%%%%%%%%%%%%%%%%%%%
	
	\subsection{Power-law model}
	\begin{eqnarray}
		\label{S_powerlawchi}
&&		\tilde{\chi}'(\omega)= \bigg[\omega ^{\kappa } \cos \left(\frac{\kappa\pi }{2}\right) \left(\gamma  \hat{\gamma}_\text{a}^2 \omega ^{2 \kappa }+\hat{k} \hat{\gamma} _\text{a} (\hat{k}+2 K)+\gamma  \hat{k}^2\right)+\hat{\gamma} _\text{a} \omega ^{2\kappa} \left(\hat{k}	\hat{\gamma}_\text{a}+\gamma  \hat{k} \cos (\kappa\pi )+\gamma  \hat{k}+\hat{\gamma}_\text{a} K\right)\nonumber\\
		&&+\hat{k}^2 K\bigg]\bigg\{
		\hat{k}^2 K^2+ 	\gamma ^2 \hat{\gamma}_\text{a}^2 \omega ^{4 \kappa }+\big[2 \gamma \hat{\gamma}_\text{a} \hat{k}K \cos (\kappa\pi )+\left(\hat{\gamma}
		_\text{a} (\hat{k}+K)+\gamma  \hat{k}\right){}^2\big] \omega ^{2 \kappa }+2 \cos \left(\frac{\kappa\pi }{2}\right)\nonumber\\
		&& \big[\hat{\gamma}_\text{a} (\hat{k}+K)+\gamma  \hat{k}\big] \big[\hat{k}K+\gamma  \hat{\gamma}_\text{a} \omega ^{2 \kappa }\big] \omega ^{\kappa }\bigg\}^{-1},
		\\
&&		\hat{\chi}''(\omega)=\,\omega ^{\kappa } \sin \left(\frac{\kappa\pi }{2}\right) \bigg[(\gamma +\hat{\gamma}_\text{a})\hat{k}^2+2 \gamma \hat{\gamma}_\text{a}  \hat{k}\cos\left(\frac{\kappa\pi }{2}\right)\omega ^{\kappa } +\gamma\hat{\gamma}_\text{a}^2  \omega ^{2 \kappa }\bigg]\bigg\{
		\hat{k}^2 K^2+ 	\gamma ^2 \hat{\gamma}_\text{a}^2 \omega ^{4 \kappa }
		\\&&
		+\big[2 \gamma \hat{\gamma}_\text{a} \hat{k}K \cos (\kappa\pi )+(\hat{\gamma}_\text{a} (\hat{k}+K)+\gamma  \hat{k})^2\big] \omega ^{2 \kappa }+2 \cos \left(\frac{\kappa\pi }{2}\right) \big[\hat{\gamma}_\text{a} (\hat{k}+K)+\gamma  \hat{k}\big] \big[\hat{k}K+\gamma  \hat{\gamma}_\text{a} \omega ^{2 \kappa }\big] \omega ^{\kappa }\bigg\}^{-1}, \nonumber\\
		\label{powerlawcorr} 
&&		\tilde C(\omega)=\,2k_{\text B}T\omega ^{\kappa-1} \sin \left(\frac{\kappa\pi }{2}\right) \bigg[(\gamma +\hat{\gamma}_\text{a}(1+\alpha))\hat{k}^2+2 \gamma \hat{\gamma}_\text{a}  \hat{k}\cos\left(\frac{\kappa\pi }{2}\right)\omega ^{\kappa } +\gamma\hat{\gamma}_\text{a}^2  \omega ^{2 \kappa }\bigg]\bigg\{
		\hat{k}^2 K^2+ 	\gamma ^2 \hat{\gamma}_\text{a}^2 \omega ^{4 \kappa }
		\nonumber\\
		&&+\big[2 \gamma \hat{\gamma}_\text{a} \hat{k}K \cos (\kappa\pi )+(\hat{\gamma}
		_\text{a} (\hat{k}+K)+\gamma  \hat{k})^2\big] \omega ^{2 \kappa }+2 \cos \left(\frac{\kappa\pi }{2}\right) \big[\hat{\gamma}_\text{a} (\hat{k}+K)+\gamma  \hat{k}\big] \big[\hat{k}K+\gamma  \hat{\gamma}_\text{a} \omega ^{2 \kappa }\big] \omega ^{\kappa }\bigg\}^{-1},
		\nonumber\\
	\end{eqnarray}
	\vskip-5mm
	\begin{equation}
		\Xi(\omega)=\frac{\hat{\gamma}_\text{a} \hat{k}^2\alpha}{(\gamma
			+\hat{\gamma}_\text{a})\hat{k}^2 +2 \gamma\hat{\gamma}_\text{a} \hat{k}\cos \left(\frac{\kappa\pi }{2}\right)\omega ^{\kappa }+\gamma \hat{\gamma}_\text{a}^2 \omega ^{2 \kappa }}.\label{powerlawspec}
	\end{equation}

	One  can see that in the limiting case  $\tau_\text{M}\rightarrow0$, 
	the Maxwell model becomes equivalent to the Newtonian fluid model. 
	Likewise, the same limit is obtained for the
	Kelvin-Voigt model  in 
	the  limits $k_{\text{KV}}\rightarrow0$ and $k_{\text{KV,a}}\rightarrow0$
	and   for the power-law model in  the limit 
	$\kappa\rightarrow1$.

	\section{\label{sec:asymptotics} Asymptotic analysis}
	Defining $\epsilon = \omega^{\kappa}$, we  expand all functions in the low-frequency limit in powers of $\epsilon$ according to 
	\begin{eqnarray}
		\tilde\chi'(\omega)\approx\,&&\frac{1}{K}-\frac{\cos \left(\frac{\kappa\pi }{2}\right) \left(\gamma +\hat{\gamma }_\text{a}\right)}{K^2}\,\epsilon+\frac{\cos (\kappa\pi ) \big[K \hat{\gamma }_\text{a}^2+\hat{k} \left(\gamma +\hat{\gamma}_\text{a}\right)^2\big]}{ \hat{k}\,K^3}\epsilon ^2+\mathcal{O}(\epsilon^3),\\
		\tilde\chi''(\omega)\approx\,&&\frac{\left(\left(\gamma +\hat{\gamma }_\text{a}\right) \sin \left(\frac{ \kappa \pi }{2}\right)\right)
		}{K^2}\,\epsilon-\frac{\big[K \hat{\gamma }_\text{a}^2+\hat{k} \left(\gamma +\hat{\gamma }_\text{a}\right){}^2\big] \sin ( 
			\kappa \pi)}{ \hat{k}K^3} \,\epsilon ^2 +\mathcal{O}(\epsilon^3),\\
		\frac{\omega\tilde C(\omega)}{2k_\text{B}T}\approx\,&&\frac{\big[\gamma +(1+\alpha ) \hat{\gamma }_\text{a}\big] \sin \left(\frac{\kappa\pi }{2}\right) }{K^2}\,\epsilon-\frac{\left(K (1+\alpha ) \hat{\gamma }_\text{a}^2+\hat{k} \left(\gamma +\hat{\gamma }_\text{a}\right) \left(\gamma +(1+\alpha )
			\hat{\gamma }_\text{a}\right)\right) \sin (  \kappa \pi)}{\hat{k}K^3 }\,\epsilon^2+\mathcal{O}(\epsilon^3),\nonumber\\
		\\
		\Xi(\omega)\approx\,&&\frac{\hat{\gamma }_\text{a} \alpha}{\gamma +\hat{\gamma }_\text{a}}-\frac{2 \cos \left(\frac{\kappa\pi }{2}\right) \gamma\hat{\gamma }_\text{a}^2\alpha}{\hat{k} \left(\gamma
			+\hat{\gamma }_\text{a}\right)^2}\,\epsilon+\frac{ \left(\gamma +2 \gamma  \cos (\kappa\pi )-\hat{\gamma }_\text{a}\right)\gamma\hat{\gamma
			}_\text{a}^3   \alpha}{\hat{k}^2 \left(\gamma +\hat{\gamma }_\text{a}\right)^3}\,\epsilon^2+\mathcal{O}(\epsilon^3).
	\end{eqnarray}
	In the high-frequency  limit, we expand in inverse powers of $\epsilon$ according to 
	\begin{eqnarray}
		\tilde\chi'(\omega)\approx\,&&\frac{\cos \left(\frac{\kappa\pi }{2}\right)}{\gamma }\epsilon^{-1}-\frac{\cos (\kappa\pi ) \left(K+\hat{k}\right)}{\gamma ^2}\epsilon^{-2}+\mathcal{O}(\epsilon^{-3}),\\ 
		\tilde\chi''(\omega)\approx\,&& \frac{\sin \left(\frac{\kappa\pi }{2}\right)}{\gamma}\epsilon^{-1}-\frac{\left(K+\hat{k}\right) \sin (\kappa\pi )}{\gamma ^2}\epsilon^{-2}+\mathcal{O}(\epsilon^{-3}),\\
		\frac{\omega\tilde C(\omega)}{2k_\text{B}T}\approx\,&&\frac{\sin \left(\frac{\kappa\pi }{2}\right)}{\gamma}\epsilon^{-1}-\frac{\left(K+\hat{k}\right) \sin (\kappa\pi )}{\gamma ^2}\epsilon^{-2}+\mathcal{O}(\epsilon^{-3}),\\
		\Xi(\omega)\approx\,&&\frac{\alpha  \hat{k}^2}{\gamma\hat{\gamma }_\text{a}}\epsilon^{-2}+\mathcal{O}(\epsilon^{-3}).	
	\end{eqnarray} 
	
	\begin{figure*}[t!]
		\begin{center}
			\includegraphics[width=\linewidth]{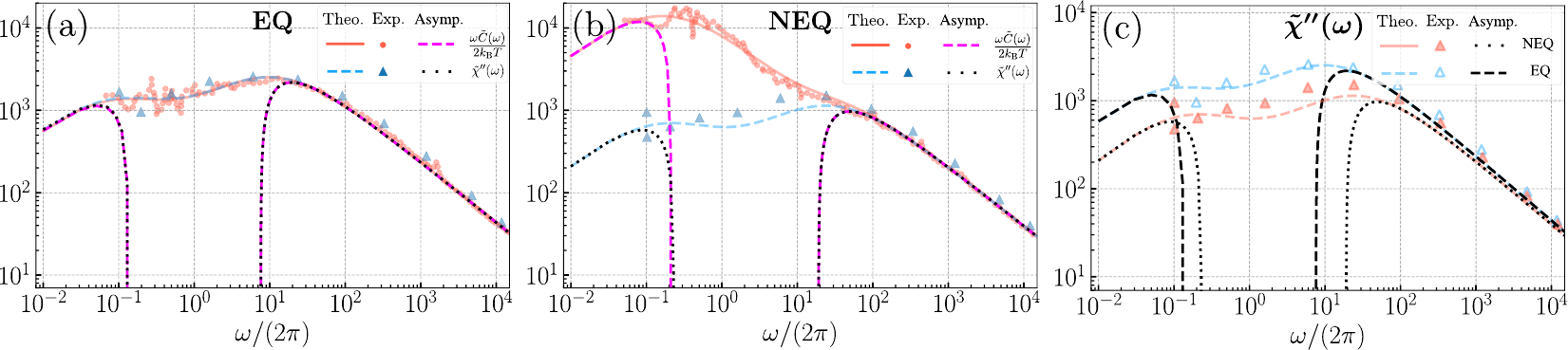}
		\end{center}
		\caption{{\em Actomyosin network: Asymptotic analysis.} The same data is shown as in  Fig.~2 in the main text, except that here
			we include the results from the asymptotic analysis. 
			(a) For the EQ case  the asymptotic expressions  for $\tilde\chi''(\omega)$ ({\em dotted black lines}) are indistinguishable
			from the asymptotic expressions  for $\omega\tilde C(\omega)/(2k_\text{B}T)$ ({\em dashed magenta lines}) 
			in both low- and high-frequency regimes. 
			(b) Results for the NEQ case. Here,  there is a clear deviation between the  asymptotic expressions
			for $\tilde\chi''(\omega)$ ({\em dotted black lines}) and $\omega\tilde C(\omega)/(2k_\text{B}T)$ ({\em dashed magenta lines}) 
			in the low-frequency regime. 
			(c) The comparison of  $\chi''(\omega)$ in the EQ and NEW cases shows the NEQ  stiffening effects.
			In all cases, the agreement between the asymptotic expressions and the full model predictions is good in the asymptotic frequency regimes. }\label{fig:asymptotic}
	\end{figure*}
	
	From these asymptotic expansions, we obtain that in the low-frequency regime,
	$\tilde{\chi}''(\omega)$ displays  a maximum located at 
	\begin{equation}
		\omega_1=\left\{\frac{K \hat{k} \left(\gamma +\hat{\gamma }_\text{a}\right)}{\big[4 K \left(\hat{\gamma }_\text{a}\right){}^2+4 \hat{k} \left(\gamma +\hat{\gamma
				}_\text{a}\right){}^2\big] \cos \left(\frac{\pi  \kappa }{2}\right)}\right\}^{1/\kappa}
	\end{equation} 
	with 
	\begin{equation}
		\tilde{\chi}''(\omega_1)=\frac{\hat{k} \left(\gamma +\hat{\gamma }_\text{a}\right){}^2 \tan \left(\frac{\pi  \kappa }{2}\right)}{8 K \big[K\hat{\gamma
			}_\text{a}^2+\hat{k} \left(\gamma +\hat{\gamma }_\text{a}\right)^2\big]}. 
	\end{equation}
	In the high-frequency regime,
	$\tilde{\chi}''(\omega)$ 
	displays a second  maximum located at 
	\begin{equation}
		\omega_2=\left\{\frac{\gamma }{4 \left(K+\hat{k}\right) \cos \left(\frac{\pi  \kappa }{2}\right)}\right\}^{1/\kappa}
	\end{equation} 
	with 
	\begin{equation}
		\tilde{\chi}''(\omega_2)=\frac{\tan \left(\frac{\pi  \kappa }{2}\right)}{8 \left(K+\hat{k}\right)}.
	\end{equation} 
	The positions of these
	maxima are in very good agreement with the experimental data, as shown in Fig.~\ref{fig:asymptotic}.
	The physical origin of these maxima is easily obtained from the analytical expressions.
	In fact,  the maximum in the low-frequency regime involves $\hat\gamma_{\text{a}}$, while the maximum in the high-frequency regime does not. 
	This shows that the low-frequency maximum is determined by the active-particle friction, while 
	the high-frequency maximum is dominated by the probe particle friction.  
	
	This is also obvious from the  frequency $\omega_1^*$, 
	at which the positional correlation function displays a maximum  in the low-frequency regime, 
	\begin{equation}
		\omega_1^*=\left\{\frac{K \hat{k} \left(\gamma +(1+\alpha ) \hat{\gamma }_\text{a}\right)}{4 \big[K (1+\alpha )\hat{\gamma }_\text{a}^2+\hat{k}
				\left(\gamma +\hat{\gamma }_\text{a}\right) [\gamma +(1+\alpha ) \hat{\gamma }_\text{a}]\big] \cos \left(\frac{\pi  \kappa }{2}\right)}\right\}^{1/\kappa}
	\end{equation} 
	with 
	\begin{equation}
		\frac{\omega_1^*\tilde C(\omega_1^*)}{2k_\text{B}T}=\frac{\hat{k} \left(\gamma +(1+\alpha ) \hat{\gamma }_\text{a}\right){}^2 \tan \left(\frac{\pi  \kappa }{2}\right)}{8 K \big[K (1+\alpha )
			\hat{\gamma }_\text{a}^2+\hat{k} \left(\gamma +\hat{\gamma }_\text{a}\right) [\gamma +(1+\alpha ) \hat{\gamma }_\text{a}]\big]}.
	\end{equation} 
	Setting $\alpha=0$ gives the same value as the low-frequency maximum of the response function,
	which implies that the difference in the position and value of the maximum of the positional autocorrelation function and the low-frequency maximum of the response function is imposed by the non-equilibrium parameter only. 
	Note that the high-frequency expression for the maximum of  $\tilde{\chi}''(\omega)$ coincides with that of  $\frac{\omega\tilde C(\omega)}{2k_\text{B}T}$. 
	But, clearly, for NEQ there is no maximum in the experimental data
	for $\frac{\omega\tilde C(\omega)}{2k_\text{B}T}$ in the high-frequency regime, 
	which implies that the first two terms of the asymptotic high-frequency expansion are 
	not sufficient for describing the positional autocorrelation function at high frequencies in the NEQ case.
	
	\section{Fitting results for different viscoelastic memory kernels}
\begin{figure*}
		\begin{center}
			\includegraphics[height=0.73\linewidth]{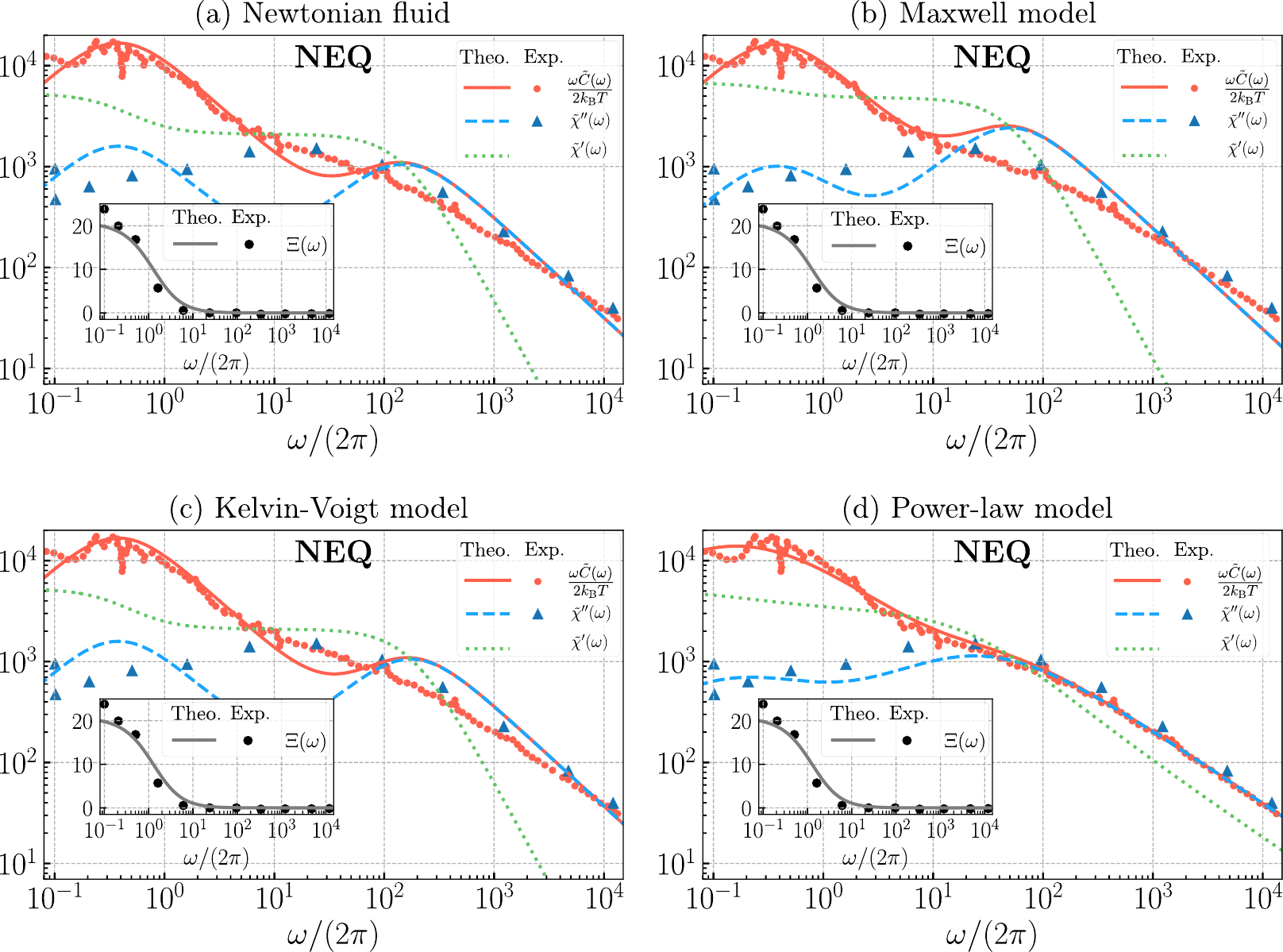}
		\end{center}  
		%	\vskip\baselineskip
		\caption{Fitting different viscoelastic models to {\em in vitro} actomyosin experiment data: (a-d) the data for the dissipative part of the response function ({\em dark blue triangles}), $\tilde{\chi}''(\omega)$, and the normalized positional autocorrelation function ({\em red circles}), $\omega \tilde{C}(\omega)/(2k_{\text{B}}T)$,  extracted directly from experimental data for F-actin network reported in Ref~\cite{Mizuno2007}, are shown. In the inset  the
			spectral function is shown ({\em black circles}), $\Xi(\omega)$. Continuous curves for $\omega \tilde{C}(\omega)/(2k_{\text{B}}T)$, $\tilde{\chi}''(\omega)$ and $\tilde{\chi}'(\omega)$ denote the model predictions, given in Sec.~\ref{supp:Allmodels}. Fitting parameters are given in Table~\ref{table:parameters-ACM}.  }
		\label{fig:comparisoninvitro}
	\end{figure*}

\begin{figure*}
		\begin{center}
			\includegraphics[height=0.73\linewidth]{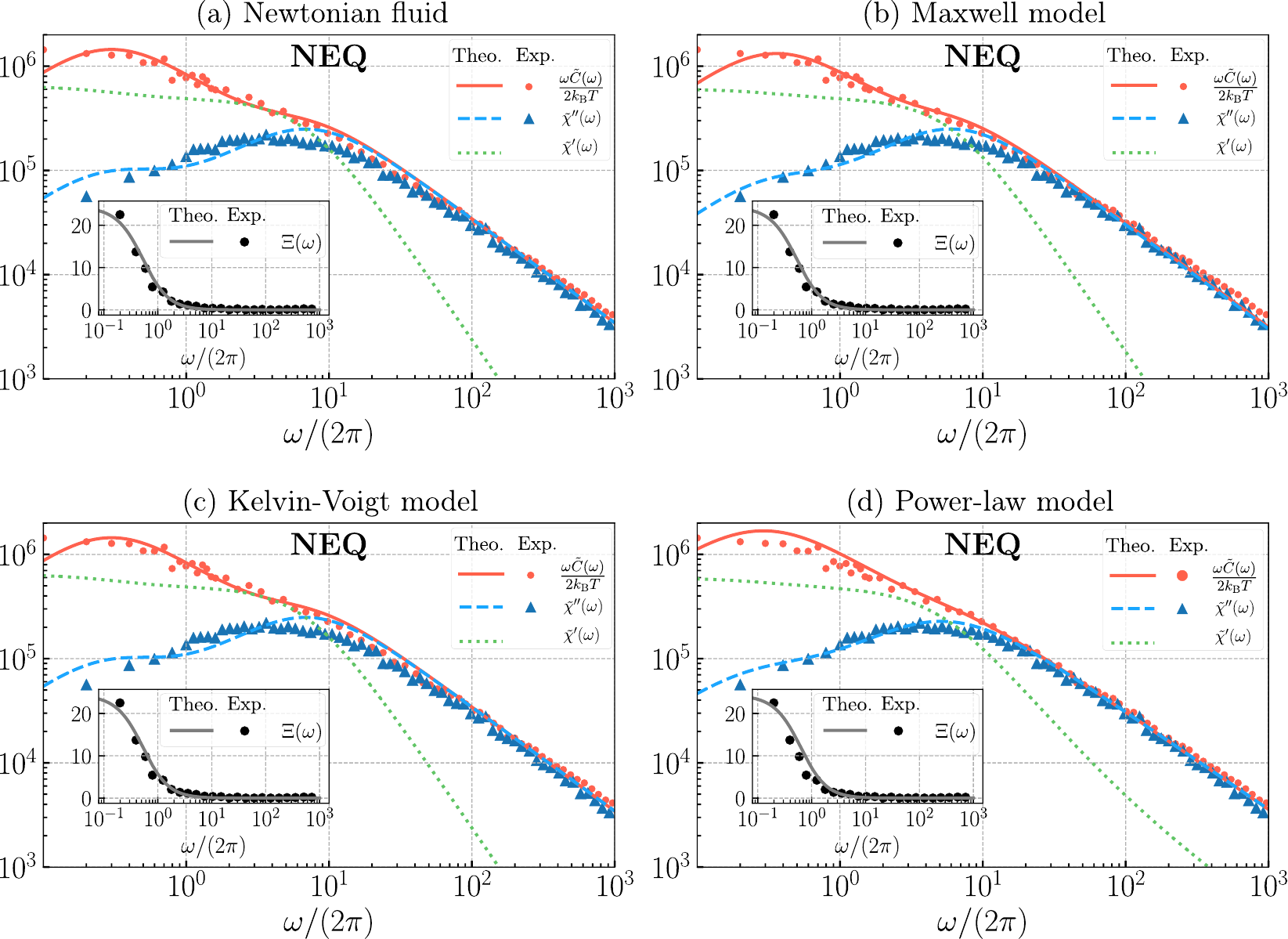}
		\end{center}  
		%\vskip\baselineskip
		\caption{Fitting different viscoelastic models to RBC flickering experiment data: (a-d) the data for the dissipative part of the response function ({\em dark blue triangles}), $\tilde{\chi}''(\omega)$, and the normalized positional autocorrelation function ({\em red circles}), $\omega \tilde{C}(\omega)/(2k_{\text{B}}T)$,  extracted directly from experimental data for F-actin network reported in Ref~\cite{Mizuno2007},  are shown. In the inset, the spectral function is shown
			({\em black circles}), $\Xi(\omega)$. Continuous curves of $\omega \tilde{C}(\omega)/(2k_{\text{B}}T)$, $\tilde{\chi}''(\omega)$ and $\tilde{\chi}'(\omega)$ denote the model predictions, given in Sec.~\ref{supp:Allmodels}. Fitting parameters are given in Table~\ref{table:parameters-RBC}.}
		\label{fig:comparisoninvivo}
	\end{figure*}

	In this section, we briefly discuss  fitting  results 
	to the NEQ experimental data (both actomyosin network and RBC flickering cases)
	for different viscoelastic memory kernels, all theoretical expressions are given in the previous section.
	For actomyosin networks (see Fig.~\ref{fig:comparisoninvitro}), none of the Newtonian, Maxwell, and Kelvin-Voigt models can capture the high-frequency tail of the experimental data, as expected, since all these models scale as $\sim\omega^{-1}$ in high-frequency limit. 
	This is different for the RBC flickering experimental data (see Fig.~\ref{fig:comparisoninvivo}), 
	where all models can fit the experimental data rather nicely. Still, the power-law model performs better than the other models.

	\begin{sidewaystable}
		\caption{\label{table:parameters-ACM} Actomyosin network experiment (NEQ case) and RBC flickering experiment (NEQ case): Fitting parameters extracted and used for plotting $\tilde{\chi}''(\omega)$, $\omega \tilde{C}(\omega)/(2k_{\text{B}}T)$ and $\Xi(\omega)$ for different models.}
		
		Actomyosin network
		\begin{ruledtabular}
			\begin{tabular}{cccccccccc}
				\textrm{Model}&
				\textrm{$\alpha$}&
				\textrm{$\gamma[\text{pN.s}/\mu\text{m}]$}&
				\textrm{$K[\text{pN}/\mu\text{m}]$}&
				\textrm{$\gamma_\text{a}[\text{pN.s}/\mu\text{m}]$}&
				\textrm{$k[\text{pN}/\mu\text{m}]$}&
				\textrm{$\tau_{\text{M}}\;[\text{s}]$}&
				\textrm{$k_{\text{KV}}[\text{pN}/\mu\text{m}]$}&
				\textrm{$k_{\text{KV,a}}[\text{pN}/\mu\text{m}]$}&
				\textrm{$\kappa$}\\
				\colrule
				Newtonian  & $9.63$ & $0.430$ & $189$ & $47.4$ & $283$ & - & - & - & -\\			
				Maxwell   & $11.5$ & $0.717$ & $196$ & $47.8$ & $213$ & $0.00$ & - & - & -\\
				Kelvin-Voigt   & $9.64$ & $0.430$ & $21.1$ & $47.4$ & $439$ & - & $12.3$ & $242$ & - \\
				
				%\hskip-1.5cm	Power-law  & $27.09$ & $5.46\times10^{-6}$ & $1.56\times10^{-4}$ & $2.98\times10^{-5}$ & $4.63\times10^{-5}$ & - & - & - & $0.74$ \\
			\end{tabular}
			
			RBC flickering
			
			\begin{tabular}{cccccccccc}\label{table:parameters-RBC}
				\textrm{Model}&
				\textrm{$\alpha$}&
				\textrm{$\gamma[\text{pN.s}/\mu\text{m}]$}&
				\textrm{$K[\text{pN}/\mu\text{m}]$}&
				\textrm{$\gamma_\text{a}[\text{pN.s}/\mu\text{m}]$}&
				\textrm{$k[\text{pN}/\mu\text{m}]$}&
				\textrm{$\tau_{\text{M}}\;[\text{s}]$}&
				\textrm{$k_{\text{KV}}[\text{pN}/\mu\text{m}]$}&
				\textrm{$k_{\text{KV,a}}[\text{pN}/\mu\text{m}]$}&
				\textrm{$\kappa$}\\
				\colrule
				Newtonian  & $19.0$ & $0.0461$ & $1.56$ & $0.185$ & $0.451$ & - & - & - & -\\
				
				Maxwell   & $22.1$ & $0.00328$ & $0.471$ & $0.0167$ & $0.0311$ & $0.0206$ & - & - & -\\
				
				Kelvin-Voigt   & $19.0$ & $0.0461$ & $0.798$ & $1.11$ & $1.10$ & - & $0.110$ & $1.59$ & -\\
				
				%\hskip-0.75cm Power-law  & $25.04$ & $5.41\times10^{-8}$ & $1.64\times10^{-6}$ & $1.25\times10^{-7}$ & $3.34\times10^{-7}$ & - & - & - & $0.98$ \\
			\end{tabular}
		\end{ruledtabular}
	\end{sidewaystable}
	
	\begin{figure*}
		\begin{center}
			\includegraphics[height=0.6\linewidth]{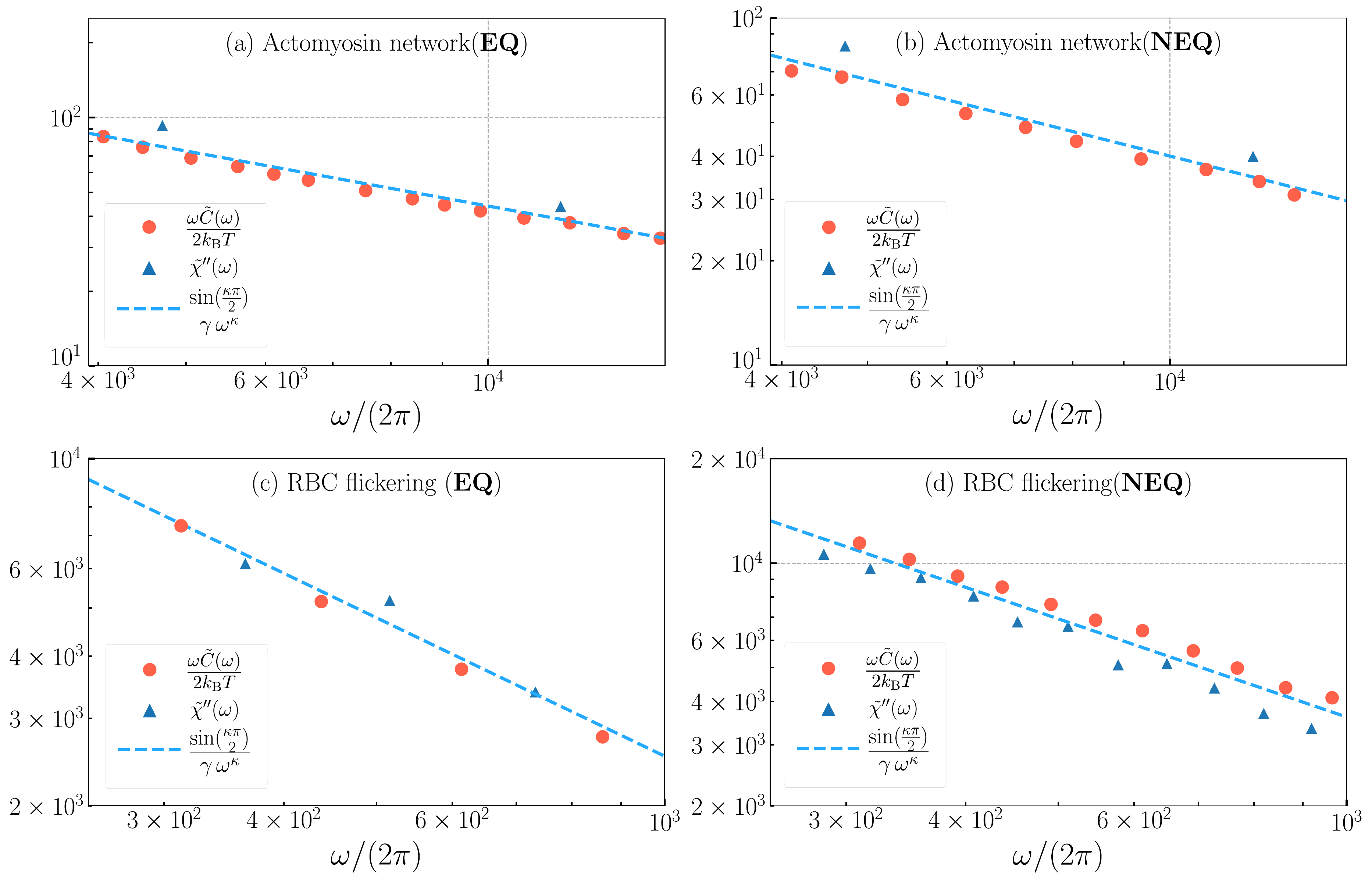}		
		\end{center}
		%\vskip\baselineskip
		\caption{Fitting the high-frequency limit of the imaginary part of the response function and rescaled correlation function in the {\em log-log} plane. 
			Filled circles and triangles are digitized experimental data for $\tilde{\chi}''(\omega)$ and $\omega \tilde{C}(\omega)/(2k_{\text{B}}T)$, respectively,
			broken lines are the asymptotic fits. Fitting parameters for each case are given in the main text.}
		\label{fig:linearfits}
	\end{figure*}
	
	\section{\label{supp:fitting} Fitting procedure}
	\subsection{Fitting models to experimental data}
	
	For both experiments discussed in this paper, we have digitized the EQ and NEQ data from Refs.~\cite{Mizuno2007,Turlier2016equilibrium} using {\em WebPlotDigitizer}\cite{Rohatgi2020}. All fits have been performed in the log-log plane. To determine $\kappa$ and $\gamma$ in the power-law model and $\gamma$ in the Newtonian fluid and Kelvin-Voigt models, we use the high-frequency limit expression of the dissipative part of the response function, given by
	$\tilde{\chi}''_\text{high-freq}(\omega) = 1/(\gamma\,\omega)$ for the Newtonian model, 
	$\tilde{\chi}''_\text{high-freq}(\omega) = 1/(\gamma\,\omega)$ for the Kelvin-Voigt model, 
	$\tilde{\chi}''_\text{high-freq}(\omega)=\frac{\gamma_\text{a}k^2\tau_\text{M}^2+\gamma(\gamma_\text{a}+k\tau_\text{M})^2}{\big[\gamma(\gamma_{\text{a}}+k\tau_\text{M})+\tau_\text{M}\big((k+K)\gamma_{\text{a}}+kK\tau_\text{M}\big)\big]^2\omega}$ for the Maxwell model and 
	$\tilde{\chi}''_\text{high-freq}(\omega) = \frac{\sin(\frac{\kappa\pi}{2})}{\gamma\,\omega^{\kappa}}$ for the power-law model,
	and fit them to the experimental  $\tilde{\chi}''(\omega)$ data in the high-frequency limit, see Fig.~\ref{fig:linearfits}. 
	For this, we use the data above  $\omega/2\pi\sim 300$ and $\omega/2\pi\sim 4000$  from the RBC flickering and actomyosin network experiments, respectively. 
	The fitted parameters are given in Table~\ref{table:parameters-ACM} and Table~\ref{table:parameters-RBC}. For the Newtonian and Kelvin-Voigt models, the slope of the high-frequency linear relation (in the log-log plane) is $-1$, and the only parameter that one needs to fit is $\gamma$. 
	For the Maxwell model, we don't fit high-frequency data separately, and we just theoretical expressions for 
	$\tilde\chi''(\omega)$ and $\omega\tilde{C}(\omega)/(2k_\text{B}T)$ simultaneously to the experimental data in the entire frequency range to get all parameters in the model. 
	After this, we fit the  theoretical expressions for 
	$\tilde\chi''(\omega)$ and $\omega\tilde{C}(\omega)/(2k_\text{B}T)$ simultaneously to the experimental data in the entire frequency range to determine the values of the remaining parameters. For the nonlinear fitting, we use MultiNonlinearModelFit from {\em Mathematica} with "NMinimize" function that employs
	a differential evolution algorithm.
	
\begin{figure*}
		\begin{center}
			\includegraphics[height=1.2\linewidth]{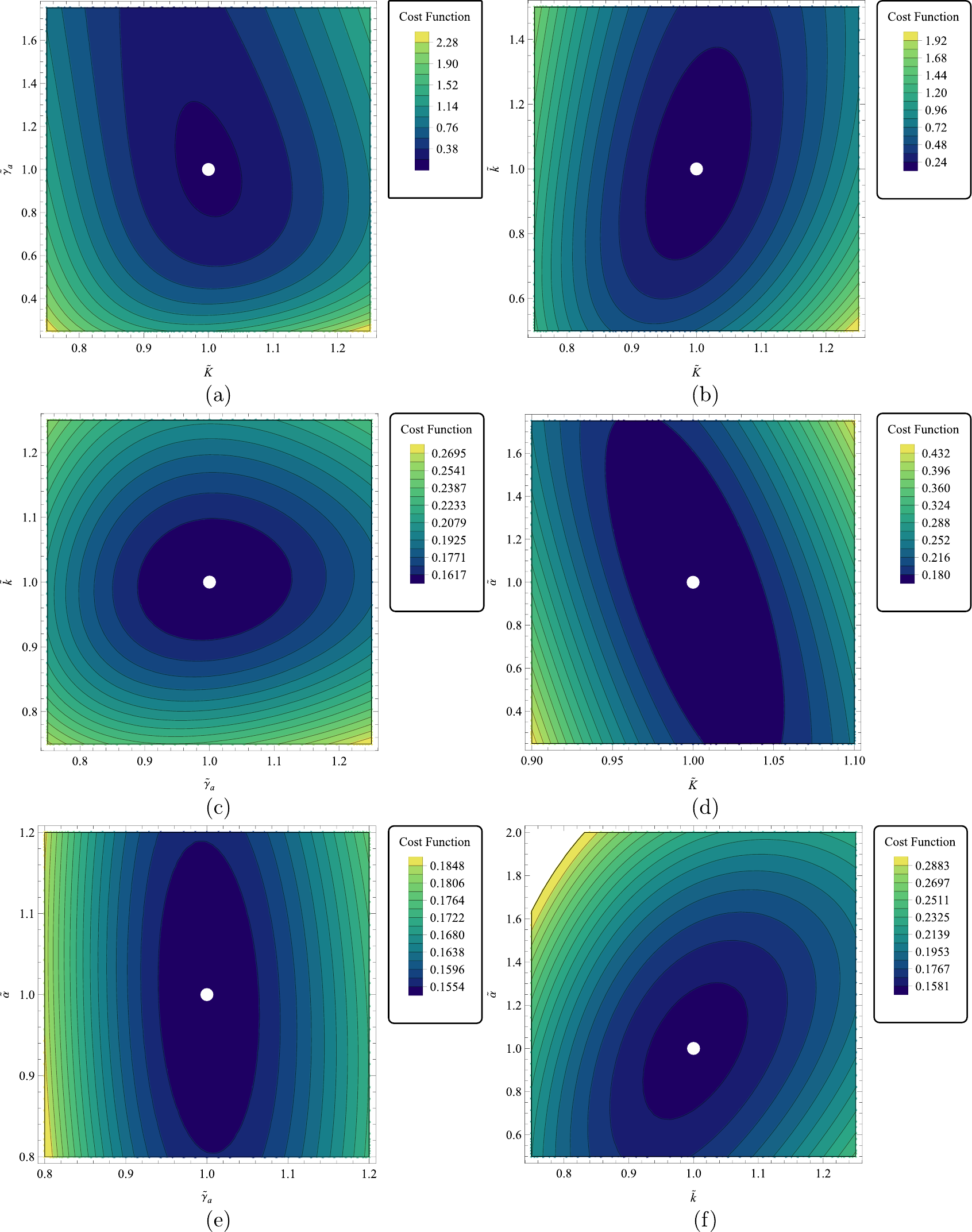}
		\end{center}
		\vskip-4mm
		\caption{Actomyosin network in EQ (a-f):  contour plots of the cost function
			Eq. \ref{eq:cost_function} 
			versus different fitting parameters. Tilde symbols denote rescaled quantities with respect to the optimal values extracted from the fitting procedure. White filled circles show the fitting parameter values obtained by our fitting procedure.}
		\label{fig:cost_function}
	\end{figure*}

	To  ensure that the fitting procedure produces a unique fit, we define
	the cost function as
	\begin{equation}
		E(\boldsymbol{x}) = \bigg[\sum_{i=1}^{N_1}(\tilde\chi''_\text{model}(\omega_i;\boldsymbol{x})-\tilde\chi''_{\text{exp.},i})^2+\sum_{i=1}^{N_2}\frac{\omega_i}{2k_\text{B}T}(\tilde C_\text{model}(\omega_i;\boldsymbol{x})-\tilde C_{\text{exp.},i})^2\bigg],
		\label{eq:cost_function}
	\end{equation}
	where $\boldsymbol{x}=(K,k,\gamma,\gamma_\text{a},\kappa,\alpha)$ is the vector of all
	model parameters and $N_1$ and $N_2$ are the number of experimental data points  for $\tilde\chi''(\omega)$ and $\omega\tilde{C}(\omega)/(2k_\text{B}T)$, respectively.
	In  Fig.~\ref{fig:cost_function} we plot the cost function Eq. \ref{eq:cost_function} in different projected parameter planes 
	for the fitting of the power-law model to actomyosin network EQ data. It is seen that the cost function 
	has a unique minimum that is correctly found by the fitting procedure.
	
	\subsection{\label{subsec:error} Error calculation}
	We assume the experimental data to obey a normal distribution with variance $\sigma^2$, which we obtain by a fit to the distribution
	of the deviation between our model prediction and the input data. 
	The error of the  parameters then follows from error propagation as the square root of the diagonal elements
	of the matrix $\sigma^2 H^{-1}$,  where 
	the Hessian matrix is defined  as $H_{ij} =\frac{\partial^2 E}{\partial x_i \partial x_j}$
	and is calculated  at the minimum of the cost function Eq.~\eqref{eq:cost_function} \cite{Press2007numerical}.
	
	%\nocite{*}
	
	\bibliography{biblio.bib}